\documentclass[letterpaper,11pt,oneside,reqno]{amsart}
\usepackage{amsfonts,amsmath, amssymb,amsthm,amscd}
\usepackage[usenames,dvipsnames]{color}
\usepackage{bm}
\usepackage{mathrsfs}
\usepackage{yfonts}
\usepackage[mpexclude,DIV13]{typearea}
\usepackage{verbatim}
\usepackage{hyperref}
\usepackage{graphicx}
\usepackage[latin1]{inputenc}
\usepackage{latexsym}
\usepackage{lscape}
\usepackage{epsfig}
\usepackage{subfigure}

\numberwithin{equation}{subsection}

\usepackage{setspace}
\include{bibtex}

\begin{document}

\bibliographystyle{alpha}
\newcommand{\e}[0]{\epsilon}
\newcommand{\EE}{\ensuremath{\mathbb{E}}}
\newcommand{\PP}{\ensuremath{\mathbb{P}}}
\newcommand{\R}{\ensuremath{\mathbb{R}}}
\newcommand{\Rplus}{\ensuremath{\mathbb{R}_{+}}}
\newcommand{\C}{\ensuremath{\mathbb{C}}}
\newcommand{\Z}{\ensuremath{\mathbb{Z}}}
\newcommand{\N}{\ensuremath{\mathbb{N}}}
\newcommand{\Q}{\ensuremath{\mathbb{Q}}}
\newcommand{\MM}{\ensuremath{\mathbb{M}}}
\newcommand{\Real}{\ensuremath{\mathrm{Re}}}
\newcommand{\Imag}{\ensuremath{\mathrm{Im}}}
\newcommand{\re}{\ensuremath{\mathrm{Re}}}
\def \Ai {{\rm Ai}}
\def \sgn {{\rm sgn}}
\newcommand{\var}{{\rm var}}
\newcommand{\I}{{\rm i}}
\renewcommand{\i}{\mathbf i}

\newtheorem{theorem}{Theorem}[section]
\newtheorem{partialtheorem}{Partial Theorem}[section]
\newtheorem{conj}[theorem]{Conjecture}
\newtheorem{lemma}[theorem]{Lemma}
\newtheorem{proposition}[theorem]{Proposition}
\newtheorem{corollary}[theorem]{Corollary}
\newtheorem{claim}[theorem]{Claim}
\newtheorem{formal}[theorem]{Critical point derivation}
\newtheorem{experiment}[theorem]{Experimental Result}
\newtheorem{prop}{Proposition}

\def\todo#1{\marginpar{\raggedright\footnotesize #1}}
\def\change#1{{\color{green}\todo{change}#1}}
\def\note#1{\textup{\textsf{\Large\color{blue}(#1)}}}

\theoremstyle{definition}
\newtheorem{remark}[theorem]{Remark}

\theoremstyle{definition}
\newtheorem{example}[theorem]{Example}

\theoremstyle{definition}
\newtheorem{definition}[theorem]{Definition}

\theoremstyle{definition}
\newtheorem{definitions}[theorem]{Definitions}

\begin{abstract}
Integrable probability has emerged as an active area of research at the interface of probability/mathematical physics/statistical mechanics on the one hand, and representation theory/integrable systems on the other. Informally, integrable probabilistic systems have two properties:
\begin{enumerate}
\item It is possible to write down concise and exact formulas for expectations of a variety of interesting observables (or functions) of the system.
\item Asymptotics of the system and associated exact formulas provide access to exact descriptions of the properties and statistics of large universality classes and universal scaling limits for disordered systems.
\end{enumerate}
We focus here on examples of integrable probabilistic systems related to the Kardar-Parisi-Zhang (KPZ) universality class and explain how their integrability stems from connections with symmetric function theory and quantum integrable systems.
\end{abstract}

\title[Macdonald processes, quantum integrable systems and the KPZ class]{Macdonald processes, quantum integrable systems and the Kardar-Parisi-Zhang universality class}

\author[I. Corwin]{Ivan Corwin}
\address{I. Corwin, Columbia University,
Department of Mathematics,
2990 Broadway,
New York, NY 10027, USA,
and Clay Mathematics Institute, 10 Memorial Blvd. Suite 902, Providence, RI 02903, USA,
and Massachusetts Institute of Technology,
Department of Mathematics,
77 Massachusetts Avenue, Cambridge, MA 02139-4307, USA
and Institut Henri Poincar\'{e},
11 Rue Pierre et Marie Curie, 75005 Paris, France}
\email{ivan.corwin@gmail.com}

\maketitle

\hypersetup{linktocpage}

%

\section{Integrable probabilistic systems in the KPZ class}

A primary aim of statistical mechanics and probability theory is to describe aggregate behavior of disordered microscopic systems driven by noise. Many systems include self-averaging mechanisms which result in the appearance of deterministic (law of large number) behavior on macroscopic scales. A central problem is to characterize the behavior of such systems between microscopic disorder and macroscopic order. On critical mesoscopic scales, large classes of systems seem to share universal fluctuation behaviors. This belief in ``universality classes'' is bolstered by (non-rigorous) physical arguments, extensive numerics, some experimental results and, recently, a growing body of mathematical proof coming from the field of integrable probability. ``Integrable'' or ``exactly solvable'' models play a key role in probing the nature and extent of universality classes. Due to enhanced algebraic structure they are often amenable to detailed analysis, thus providing the most complete access to various phenomena such as phase transition, scaling exponents, and fluctuation statistics.

The success of integrable probability in describing universal behaviors is quite strking for the non-equilibrium statistical mechanics problem of describing random interface growth. In this section we provide examples of integrable probabilistic systems whose analysis deepens our understanding of random $(1+1)$-dimensional random interface growth and the Kardar-Parisi-Zhang (KPZ) universality class. Through these examples we also demonstrate connections to interacting particle systems (models for traffic flow, queuing, mass transport, driven gases, and shock-fronts), directed polymers in random media (models for competition interfaces, domain walls, and cracking interfaces), and parabolic Anderson models (models for population growth with migration). See the review \cite{ICReview} for further background and references.

The study of KPZ universality was initiated by Kardar-Parisi-Zhang \cite{KPZ} in 1986 and drew heavily on earlier work of Forster-Nelson-Stephens \cite{FNS} in 1977. The ensuing decade of physical theories, numerics, and experiments produced strong physical evidence for universality of random interface fluctuations in their long-time and large-scale limits. The fluctuation scaling exponent and transversal correlation length was predicted to be $1/3$ and $2/3$ (meaning fluctuations of order $t^{1/3}$ correlated over distances $t^{2/3}$, with $t$ measuring time). Certain statistics (e.g. skewness, kurtosis, tail decay) were also computed numerically during this period.

The involvement of mathematicians and the first (mathematically) rigorous results and exact formulas for fluctuation statistics came in 1999 with the work of Baik-Deift-Johansson \cite{BDJ} and Johansson \cite{KJ}. Our first example below details some of Johansson's \cite{KJ} asymptotic analysis results on TASEP. The methods of determinant point process or Schur measure / process (of which TASEP is a special limiting case) have driven many further advances in understanding KPZ class statistics (see \cite{OkResh,BF,BorGorreview,BorProc}). All those models analyzed by these determinantal methods are ``totally asymmetric'' or ``zero temperature''.

The first analysis (to the point of asymptotic statistics) of a non-determinantal KPZ class model was performed by Tracy and Widom \cite{TW1,TW2,TW3} in 2009. Since then, a variety of methods have been developed to discover and analyze non-determinantal ``partially asymmetric'' or ``positive temperature'' models. 

In this paper we describe some facets of the exact solvability of $q$-TASEP, the O'Connell-Yor semi-discrete directed polymer, ASEP, and the KPZ equation. We then develop two methods used in studying these examples: (1) the theory Macdonald processes, which is an algebraic framework for discovering and analyzing a variety of probabilistic system by leveraging the remarkable properties of Macdonald symmetric polynomials; (2) the theory of quantum integrable systems, which is based on the (coordinate / algebraic) Bethe ansatz and provides a means to diagonalize certain Hamiltonians, including some stochastic generators related to the processes with which we are concern. We develop both of these methods at a high combinatorial (or algebraic) level and thus avoid many of the analytic issues and demystify the apparent algebraic miracles which arise in various degenerations. For instance, our treatment of $q$-TASEP in Section \ref{SECQIS} can be considered a mathematically rigorous version of the replica method for directed polymers \cite{K,Dot,CDR}.

There are many other exciting recent developments related to the KPZ which we will not discuss at any length. To name a few, these include: tropical combinatorics and directed polymers \cite{OCon,COSZ,OSZ}, line ensembles \cite{CH,CH2,OConWar}, coupling methods and second class particles \cite{BQS,SeppLog,BKS}, spectral methods \cite{QaustelValkoASEP,QaustelValkoLogGas}, experiments confirming KPZ statistics \cite{TS,TSSS,Coffee}, KPZ equation well-posedness \cite{Hairer, Hairer2}.

\subsection{Example 1: TASEP}

\begin{figure}
\includegraphics[height=5cm]{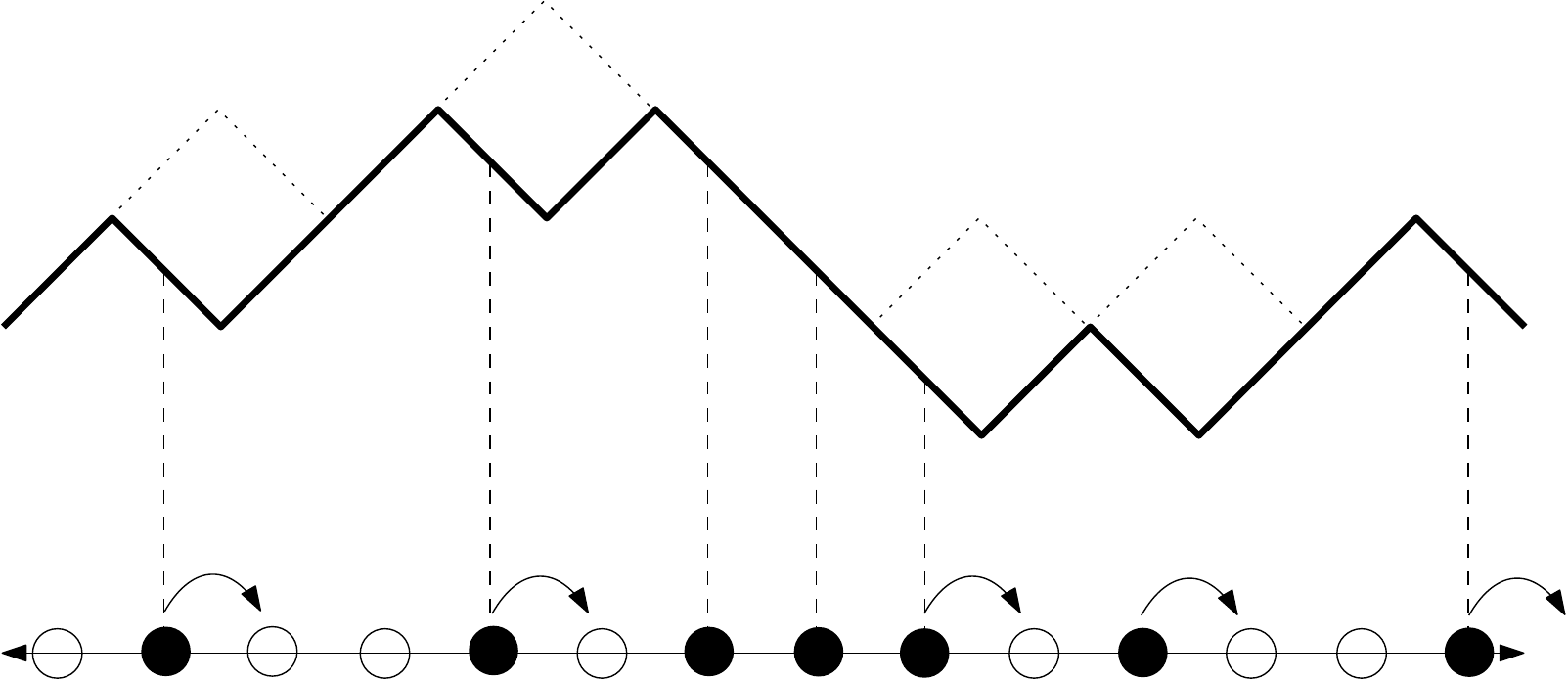}
\caption{TASEP particle configuration and height function with possible right jumps of particles denoted by arrows and possible height function growth locations denoted by dotted wedges.}
\label{TASEP}
\end{figure}

The totally asymmetric simple exclusion process (TASEP) is an interacting particle system on $\Z$. Particles inhabit sites of $\Z$ with only one particle per site at any given time. In continuous time, particles attempt to orchestrate independent random jumps according to rate one exponential clocks (in other words, according to exponential distributed waiting times of rate one) by one site to the right. If the destination site is occupied, the jump is suppressed. This process may be described in terms of occupation variables which track of whether sites of $\Z$ are occupied, or particle location variables which tracks the location of indexed particles. We will, instead, appeal to a ``height function'' to describe this process. The TASEP height function is a piece-wise linear function made up of unit $+1$ or $-1$ slope line increments. Above every site of $\Z$ with a particle, there is a $-1$ slope and above every site without a particle there is a $+1$ sloped. The height function $h^{{\rm TASEP}}(t,x)$ pastes these increments together into a continuous function (uniquely defined up to an overall height shift). TASEP dynamics corresponds to replacing local minima $\vee$ by maxima $\wedge$ according to rate one exponential clocks. In this language of height functions, TASEP is equivalent to the {\it corner growth model}. This is illustrated in Figure \ref{TASEP}.

Johansson \cite{KJ} computed an exact formula for the one-point distribution of the TASEP height function, when initialized from ``step'' or ``wedge'' initial data. This initial data corresponds to starting with every site to the left of the origin occupied, and all other sites empty. In terms of the height function this corresponds to starting with $h^{{\rm TASEP}}(0,x)=|x|$.

\subsubsection{KPZ class asymptotics}
Studying the long-time, large-scale fluctuation behavior of this height function revealed the first exact formulas for statistics of the KPZ universality class. The prediction coming from \cite{KPZ,FNS} was that in large time $L$, the height function should be non-trivially correlated on a scale of order $L^{2/3}$ with fluctuations of order $L^{1/3}$. Define the scaled height function
$$
h^{{\rm TASEP}}_{L}(t,x) := L^{-1/3} \Big(h^{{\rm TASEP}}(Lt, L^{2/3} x) - \frac{Lt}{2}\Big),
$$
where $L$ is a large scaling parameter and the centering by $Lt/2$ follows from the hydrodynamic theory for TASEP. Johansson \cite{KJ} showed the following:
\begin{theorem}\label{THMTASEPGUE}
For TASEP with step initial data
$$
\lim_{L\to \infty} \PP\left(h^{{\rm TASEP}}_{L}(1,0) \geq -s\right) = F_{{\rm GUE}}(s)
$$
where $F_{{\rm GUE}}(s)$ is the Tracy-Widom limit distribution \cite{TW} for the largest eigenvalue of a large Hermitian random matrix.
\end{theorem}
This result provides an exact prediction for the limiting one-point behavior of a wide class of models which share general characteristics with TASEP (and which are started from step type initial data). In many ways, this $F_{{\rm GUE}}$ distribution is to $(1+1$)-dimensional random growth as the Gaussian distribution is to random walks. Asymptotic analysis of the remaining examples we discuss yield the same scaling and distributional limit, thus providing further evidence for KPZ universality.

Using methods of determinantal point processes, further exact statistics describing the KPZ class has been extracted through studying the large $L$ limit of $h^{{\rm TASEP}}_{L}(t,x)$ (e.g. the Airy processes describing the  fixed  $t$ and varying $x$ limit for step and a few other types of initial data, see \cite{ICReview}). The connection between random matrix theory and growth processes will be alluded to further in our discussion of Macdonald processes in Section \ref{SECMAC}. TASEP is one of a handful of examples of particle systems and growth models which are analyzable via determinantal point processes (or equivalently Schur processes, free Fermions, non-intersecting paths). These other examples are discrete time TASEPs with sequential or parallel update rules, pushASEP or long range TASEP, directed last passage percolation in two dimension with geometric, exponential or Bernoulli weights, and the polynuclear growth process -- see \cite{BorGorreview,BorProc} and references therein.

\medskip

The examples which we address here are deformations (and limits of deformations) of TASEP. These examples are no longer determinantal, though there still turn out to be large families of observables whose averages are explicit (for determinantal systems correlation functions are written explicitly as determinants). Integrability is quite sensitive to perturbations and while these deformations are integrable,  there are many simple models, closely related to TASEP, which are not.

\subsection{Example 2: $q$-TASEP}

\begin{figure}
\includegraphics[width=12cm]{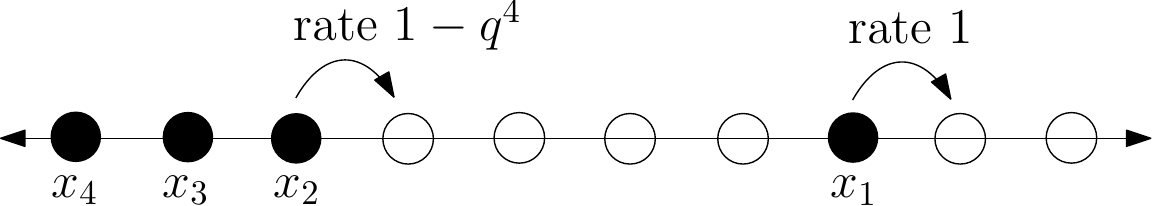}
\caption{In $q$-TASEP particles jump rate according to rate $1-q^{{\rm gap}}$ exponential clocks where gap is the distance to the next particle.}
\label{qTASEP}
\end{figure}

The $q$-deformed totally asymmetric simple exclusion process ($q$-TASEP) is a one parameter deformation of TASEP which was discovered and first studied in the context of Macdonald processes \cite{BorCor} (see also subsequent work \cite{BCS,BCdiscrete,BCF,FerVet,BCPS,OConPei,Cqmunu,Marko}). Fix $q\in (0,1)$ and let $x_i(t)\in \Z$ be the location of particle $i$ at time $t$. We assume that $x_j(t)<x_{i}(t)$ for $j>i$ and that there is a right-most particle which we label $x_1(t)$ (for notational convenience fix $x_0(t)\equiv +\infty$). In continuous time, each particle $x_i$ attempts to jump one site to the right according to an exponential clock of rate $1-q^{x_{i-1}(t)-x_i(t)-1}$. Here $x_{i-1}(t)-x_{i}(t)-1$ is the number of empty sites between particle $i$ and the next-right particle $i-1$. This jump rate interpolates between rate zero when the gap is zero and rate one when the gap tends to infinity. This can be thought of as a traffic model in which cars (particles) slow down as they approach the car in front of them. A value of $q$ near one represents a road with cautious drivers. When $q$ goes to zero TASEP is recovered and cars move without caution, only yielding immediately before an accident (when two particles would occupy the same site). This process is illustrated in Figure \ref{qTASEP}. Note that $x_1$ always jumps to the right at rate one since the distance to $x_0\equiv +\infty$ is infinite.

\subsubsection{Moment formulas}
The integrability of $q$-TASEP is partially captured in the following theorem, initially proved when all $n_i\equiv n$ in \cite{BorCor}, and then for general $n_i$ in \cite{BCS,BCGS}. Sections \ref{SECMAC} and \ref{SECQIS} describe the two methods for proving this theorem.

\begin{figure}
\includegraphics[height=5cm]{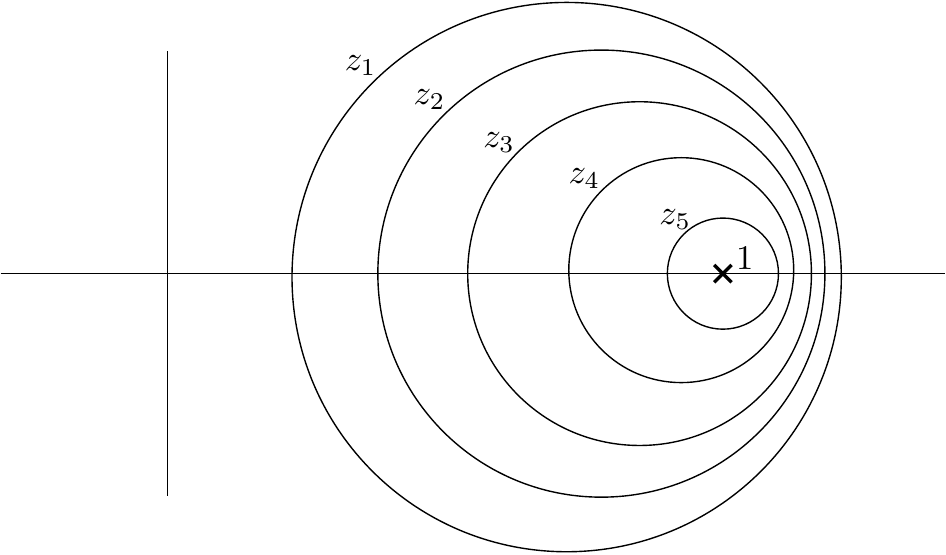}
\caption{Possible contours for Theorem \ref{THMqTASEPmoments} when $k=5$ and $q$ is close to 1.}
\label{qContours}
\end{figure}

\begin{theorem}\label{THMqTASEPmoments}
Consider $q$-TASEP with step initial data ($x_n(0)=-n$, $n\geq 1$). For all $k\geq 1$ and $n_1\geq n_2\geq \cdots \geq n_k\geq 1$,
\begin{equation}\label{EQNqTASEPmoments}
\EE\bigg[\prod_{j=1}^{k} q^{x_{n_j}(t)+ n_j}\bigg] = \frac{(-1)^k q^{\frac{k(k-1)}{2}}}{(2\pi \I)^k} \oint \cdots \oint \prod_{1\leq A<B\leq k} \frac{z_A-z_B}{z_A-qz_B} \prod_{j=1}^{k} \frac{ e^{(q-1)tz_j}}{(1-z_j)^{n_j}} \frac{dz_j}{z_j},
\end{equation}
where, for each $A\in \{1,\ldots,k\}$ the contour of integration of $z_A$ contains $1$, as well as $q$ times the contour of integration of $z_B$ for $B>A$, but does not contain $0$ (see Figure \ref{qContours}).
\end{theorem}

Step initial data means $x_n(t)+n= 0$ for all $n\geq 1$, and since particles only move to the right this implies that the random variables $q^{x_n(t)+n}$ are in $(0,1]$ for all $t\geq 0$. The knowledge of all joint moments uniquely identifies the joint distributions of this collection of random variables, and hence that of all $x_n(t)$ for fixed $t$ and varying $n$. The challenge is to extract exact distributional formulas from the results of Theorem \ref{THMqTASEPmoments} in such that they are amenable to asymptotic analysis. So far, this has only been successfully implemented for the one-point distribution (i.e. distribution of $x_n(t)$ for a fixed $n$ and $t$), as we now describe.

\subsubsection{Fredholm determinant}\label{SECFREDDETs}

Theorem \ref{THMqTASEPmoments} with all $n_k\equiv n\geq 1$ yields
$$
\EE\Big[q^{k(x_{n}(t)+ n)}\Big] = \frac{(-1)^k q^{\frac{k(k-1)}{2}}}{(2\pi \I)^k} \oint \cdots \oint \prod_{1\leq A<B\leq k} \frac{z_A-z_B}{z_A-qz_B} \prod_{j=1}^{k} \frac{ e^{(q-1)tz_j}}{(1-z_j)^{n}} \frac{dz_j}{z_j},
$$
with nested contours as in Theorem \ref{THMqTASEPmoments}. Nested integrals become cumbersome as $k$ grows (we will need to utilize these formulas for all $k\geq 1$), so it is natural to deform our formulas so the contours remain fixed as $k$ varies. Such contour deformations can be made, though they necessarily involve deforming through poles. By keeping track of the residues from crossing these poles, the complexity of the nested contours is transferred into complexity of the integrand.

There are two ways to ``un-nest'' the contours. One way is to deform them sequentially ($z_1$ through $z_k$) to lie upon a large contour containing both 0 and 1. This deformation crosses the simple pole of the integrand at $z_j=0$ for all $j$. The other way is to deform them sequentially ($z_{k}$ through $z_1$) to lie upon a small contour containing only 1. This deformation crosses the simple poles of the integrand coming from the denominator $z_A-q z_B$. Both deformations ultimately yield formulas for the distribution of $x_n(t)$. Though the second deformation is slightly more involved, it also yields a formula which ends up being more readily amenable to asymptotic analysis. We do not detail these residue considerations as they are explained at length in \cite{BorCor,BCS,BCPS}.

Consider the moment generating function
$$
\sum_{k=0}^{\infty} \EE\Big[q^{k(x_{n}(t)+ n)}\Big] \frac{\zeta^k}{(1-q)\cdots (1-q^k)}.
$$
For $|\zeta|$ small enough this is convergent. Using the second deformation described above, and carefully keeping track of the residues which arise, this generating function is rewritten as the Fredholm determinant expansion
\begin{align*}
\sum_{k=0}^{\infty} \EE\Big[q^{k(x_{n}(t)+ n)}\Big] \frac{\zeta^k}{(1-q)\cdots (1-q^k)} &= \det\big(I+K^{q-TASEP}\big)_{L^2(C_1)}\\
 &:= 1+ \sum_{L=1}^{\infty} \frac{1}{L!} \oint_{C_1}\frac{dw_1}{2\pi \I} \cdots  \oint_{C_1}\frac{dw_1}{2\pi \I} \det\big(K^{q-TASEP}(w_i,w_j)\big)_{i,j=1}^{L}
\end{align*}
where $C_1$ is a small circle around 1, and
$$
K^{q-TASEP}(w,w') = \sum_{n=1}^{\infty }\frac{f(w)\cdots f(q^{n-1}w)}{q^n w- w'}, \qquad\textrm{with} \quad f(w) = e^{(q-1) t w}{(1-w)^{-n}}.
$$
The summation in $K^{q-TASEP}$ can be replaced by a ``Mellin-Barnes'' contour integral as
\begin{equation}\label{eqnK}
K^{q-TASEP}(w,w') = \frac{1}{2\pi \I} \int_{1/2 - \I\infty}^{1/2+\I\infty} \Gamma(-s)\Gamma(1+s) (-\zeta)^s e^{-(1-q^s)tw} \left(\frac{(w;q)_{\infty}}{(q^sw;q)_{\infty}}\right)^n\, \frac{1}{q^s w-w'} \,ds.
\end{equation}
This replacement of summation by integral is important for studying asymptotics (such as those which yield Theorem \ref{THMOYLaplace}). Though the summation formula for $K^{q-TASEP}$ becomes highly oscillatory (with no termwise limit), the integral involves a contour in which the integrand has clear and well-controlled asymptotic behavior.

Owing to the fact that $q^{x_n(t)+n}\in (0,1]$, for $|\zeta|$ small enough we may interchange the expectation and the infinite summation over $k$ so that
$$
\EE\bigg[\sum_{k=0}^{\infty} q^{k(x_{n}(t)+ n)} \frac{\zeta^k}{(1-q)\cdots (1-q^k)}\bigg] = \det\big(I+K^{q-TASEP}\big)_{L^2(C_1)}.
$$
An application of the $q$-Binomial theorem simplifies the left-hand side yielding:
\begin{theorem}\label{THMqTASEPFRED}
Consider $q$-TASEP with step initial data. For any $n\geq 1$, $t\geq 0$ and $\zeta\in \C\setminus \R_{+}$
\begin{equation*}
\EE\bigg[\frac{1}{(\zeta q^{x_{n}(t)+ n};q)_{\infty}}\bigg] = \det\big(I+K^{q-TASEP}\big)_{L^2(C_1)},
\end{equation*}
where $(a;q)_{\infty} := (1-a)(1-qa)(1-q^2 a)\cdots$ is the $q$-Pochhammer symbol, the operator $K^{q-TASEP}$ is given by (\ref{eqnK}) and the contour $C_1$ is a small circle around $1$.
\end{theorem}
The expression on the left is known as the $e_{q}$-Laplace transform of $q^{x_n(t)+n}$ and dates back to 1949 work of Hahn \cite{Hahn}. Like the Laplace transform of a positive random variable, this $e_q$-Laplace transform can be inverted to compute the distribution of the random variable \cite[Proposition 3.1.1]{BorCor}.

\subsubsection{KPZ class asymptotics}

Asymptotic analysis of this formula yields a generalization of Theorem \ref{THMTASEPGUE} (which corresponds with $q=0$). This was performed by Ferrari-Vet\H{o} \cite{FerVet} who showed that for any $c>c^*$ ($c^*$ should be 0, though \cite{FerVet} deal with some strictly positive value as it simplifies aspects of the analysis), and suitable $c'=c'(c,q)>0$ and $c''=c''(c,q)>0$, as $L\to \infty$, $c'' L^{-1/3} \big(x_{cL}(L)- c'L\big)$ converges in distribution to $F_{{\rm GUE}}$ (just as in the case of TASEP). This demonstrates how the $L^{1/3}$ scaling and $F_{{\rm GUE}}$ limit theorem is not unique to TASEP, but rather extends to the whole family of $q$-TASEPs.
%

\subsection{Example 3: O'Connell-Yor semi-discrete random polymer}

A $q\to 1$ limit of $q$-TASEP leads to a systems of SDEs which can be thought of as a continuous space interacting particle systems. Exponentiating this system yields a semi-discrete version of the stochastic heat equation which is a special case of the parabolic Anderson model and also describes the evolution of the partition function for the semi-discrete random polymer model introduced by O'Connell-Yor \cite{OY}.

\subsubsection{Limit of $q$-TASEP as $q\to 1$}\label{SEClimq}

Recall that $q\in (0,1)$ controls the length scale on which particles in $q$-TASEP tend to separate. Let $q = e^{-\e}$ with $\e>0$ a scaling parameter which will tend to zero. The behavior of $x_1(t)$ is quite simple. Since we have fixed $x_0(t)\equiv +\infty$, $x_1(t)$ orchestrates a simple Poisson jump process in which it increases its value by one according to an exponential rate one clock. Thus (regardless of $q$), the central limit theorem implies that
$$
\e \big(x_1(\e^{-2} \tau) -\e^{-2}\tau\big) \to B_1(\tau)
$$
where $B_1$ is a standard Brownian motion.

The behavior of $x_n(t)$ for $n>1$ requires further consideration, and different scaling:
$$
t= \e^{-2} \tau,\qquad x_n(t) = \e^{-2}\tau - (n-1)\e^{-1} \log \e^{-1} - \e^{-1} F^{\e}_n(\tau).
$$
Under this scaling it is shown in \cite[Theorem 4.1.26]{BorCor} (see also \cite[Proposition 6.2]{BCS}) that $\big\{F^{\e}_n(\cdot)\big\}_{n\geq 1}$ converges to $\{F_n\}_{n\geq 1}$ which solves the systems of SDEs
$$
d F_n(\tau)  = e^{F_{n-1}(\tau)-F_{n}(\tau)}d\tau  + dB_n(\tau)
$$
for independent Brownian motions $\{B_n\}_{n\geq 1}$ (with the convention that $F_0(\tau)\equiv -\infty$). Indeed, once $x_n$ and $x_{n-1}$ are separated by roughly $\e^{-1}\log \e^{-1}$, $x_n$ jumps ahead at rate $1-q^{x_n(\tau)-x_{n-1}(\tau)-1} \approx 1- \e e^{F^{\e}_{n-1}(\tau)-F^{\e}_{n}(\tau)}$. In time of order $\e^{-2}$, this $\e$ correction to the jump rate only affects the overall drift, thus yielding the claimed SDEs. The initial data for this system corresponding to step initial data for $q$-TASEP can either be described via an entrance law, or through an exponential transform (as now done).

Define semi-discrete stochastic heat equation (SHE) with multiplicative noise as the system of SDEs
\begin{equation}\label{eqnztaundiff}
d z(\tau,n) = \nabla z(\tau,n) + z(\tau,n) dB_n(\tau)
\end{equation}
with $(\nabla f)(n) = f(n-1)-f(n)$, and independent Brownian motions $\{B_n\}_{n\geq 1}$ (with the convention that $z(\tau,0)\equiv 0$). By It\^{o}'s lemma, $z(\tau,n) = e^{-\frac{3}{2} \tau + F_{n(\tau)}}$.
The semi-discrete SHE initial data which comes from step initial data for $q$-TASEP is $z(\tau,n) = \mathbf{1}_{n=1}$, i.e. the {\it fundamental solution}.

\subsubsection{Parabolic Anderson model}
The semi-discrete SHE in (\ref{eqnztaundiff}) arises in a simple model for population growth and migration in a random environment. Consider an ensemble of unit mass particles in $\Z$ that evolve according to the following rules:
At each time $\tau\geq 0$, and location $n\in \Z$, each resident particle
\begin{itemize}
\item splits into two identical unit mass particles, at exponential rate $r_{+}(\tau,n)$;
\item dies at an exponential rate $r_{-}(\tau,n)$;
\item jumps to the right by one at an exponential rate $1$.
\end{itemize}
The functions $r_{+}$ and $r_{-}$ represent an environment in which the particles of this system evolve. Individual particles do not feel each other (and many can occupy the same site), as the exponential clocks controlling their splits, deaths and jumps are independent. A variant of the Feynman-Kac representation implies that the expected total mass $z(\tau,n)$ satisfies
$$
\frac{d}{d\tau} z(\tau,n) = \nabla z(\tau,n) + z(\tau,n) \big(r_{+}(\tau,n)-r_{-}(\tau,n)\big).
$$
We have used $z$ here since if the media is rapidly mixing in time and space, the environment $r_{+}(\tau,n)-r_{-}(\tau,n)$ may be modeled by independent white-noises $dB_{n}(\tau)$, in which case the above equation becomes (\ref{eqnztaundiff}). The fundamental solution corresponds to starting a cluster of particles at location 1, and nowhere else.

This population model is called a parabolic Anderson model and has been extensively studied within probability literature \cite{CarMol} (see also \cite{BCLyapunov, Fest, denHoll, CC} and references therein). Since the population will generally grow/die exponentially, it is natural to study $\log z(\tau,n)$. The spikes in this function record population explosions and can be studied in terms of the phenomenon called {\it intermittency}, while the typical fluctuations correspond to a semi-discrete variant of the KPZ equation (or a continuous space interacting particle system in which Brownian motions interact in an exponential potential with the next lowest index Brownian motion). We will start by investigating the atypical behavior of $\log z(\tau,n)$, and then turn to the typical.


\subsubsection{Intermittency and Lyapunov exponents}
Systems with intermittency display large spikes distributed in time, space and magnitude in a certain multi-fractal manner. In the 1980's it was argued that such a phenomena arises in magnetic fields in turbulent flows, like those on the surface of the Sun \cite{Zel}. The idea of intermittency seems to date back at least to \cite{Bat,Kol} (see \cite{BCLyapunov,Davar,BertiniCancrini, denHoll} and references therein for more recent developments).

The mathematical definition of intermittency given in \cite{CarMol} captures a portion of this phenomenon (though not the full multi-fractal space-time structure). The $p$-th moment Lyapunov exponent $\gamma_p$ ($p\geq 1$) and almost sure Lyapunov exponent $\tilde\gamma_1$ are defined as
$$
\gamma_p(\nu) := \lim_{\tau\to \infty} \frac{1}{\tau} \log \EE\left[z(\tau,\nu\tau)^p\right],\qquad \textrm{and}\qquad
\tilde \gamma_1(\nu) :=\lim_{\tau\to \infty} \frac{1}{\tau} \log z(\tau,\nu\tau)
$$
where $\nu>0$ determines the ratio of $n/\tau$. We say that $z(\tau,n)$ displays {\it intermittency} if
$$
\tilde\gamma_1 <\gamma_1<\frac{\gamma_2}{2}<\frac{\gamma_3}{3} < \cdots.
$$
Such an ordering has a clear interpretation. That $\tilde\gamma_1<\gamma_1$ implies that the first moment of $z(\tau,n)$ is not determined by the typical behavior of $\log z(\tau,n)$, but rather by its uncommonly high peaks. In general, the growth of these moments reflects the fact that $\log z(\tau,n)$ has a sufficiently heavy upper tail so that moments are dominated by higher and higher peaks, of smaller and smaller probabilities. At a typical location $n$, these high peaks will not appear, however, over a wide range it is likely to see quite large peaks.

Under the scaling described in Section \ref{SEClimq} the formulas Theorem \ref{THMqTASEPmoments} provided for $\EE\big[q^{k(x_n(t)+n)}\big]$ converge to corresponding formulas for $\EE[z(\tau,n)^p]$ (as shown in \cite[Theorem 1.8]{BCLyapunov}).
\begin{proposition}
The moment Lyapunov exponents for the fundamental solution to the semi-discete SHE are given by
$$
\gamma_p(\nu) = H_p(z_p^0)\, \qquad \textrm{where}\qquad
H_p(z) = \frac{p(p-3)}{2} + pz - \nu \log \left(\prod_{i=0}^{p-1}(z+i)\right)
$$
and $z_p^0$ is the unique solution to $H_p'(z) = 0$  with $z\in (0,\infty)$.
\end{proposition}
The almost sure Lyapunov exponent was conjectured in \cite{OY} and proved in \cite{OConnellMoriarty} (see also Theorem \ref{THMsdSHEasy}) to be
$$
\tilde\gamma_1(\nu) = -\frac{3}{2}  + \inf_{s>0}\big(s - \nu \Psi(s)\big)
$$
where $\Psi(s) := \big[\log \Gamma\big]'(s)$ is the digamma function. These formulas demonstrate a very explicit confirmation of intermittency.

\subsubsection{O'Connell-Yor semi-discrete directed polymer in random media}

Whereas the atypical behavior of $\log z(\tau,n)$ is quite interesting through the lens of the parabolic Anderson model, it is the typical behavior which is most important when considering this as an interacting particle system or directed polymer model. The solution to the system of SDEs (\ref{eqnztaundiff}) satisfied by $z(\tau,n)$ can be written in path integral form via the Feynman-Kac representation as
\begin{align}\label{eqnztau}
z(\tau,n) &= \EE_{x(\tau)=n}\left[\mathbf{1}_{x(0)=1} \exp\left\{\int_{0}^{\tau} dB_{x(s)}(s) - \frac{\tau}{2}\right\}\right] \\
\nonumber &= e^{-\frac{3}{2} \tau} \int\displaylimits_{0<s_1<\cdots <s_{n-1}<\tau} e^{B_1(0,s_1)+\cdots + B_n(s_{n-1},\tau)} ds_1\cdots ds_{n-1}
\end{align}
where the expectation $\EE_{x(\tau)=n}$ is over Poisson jump processes (which increase value by one at exponential rate one) which are pinned to be $n$ at time $\tau$ (in other words, $x(\cdot)$ is a Poisson jump process run backwards in time from $n$ at time $\tau$, decreasing by one at rate one in backwards time). The second line follows since the trajectory of a Poisson jump process $x(\cdot)$ pinned at $x(\tau)=n$ and $x(0)=1$ is (up to a normalization by $e^{-\tau}$) distributed uniformly over the simplex of possible jumping times $0<s_1<\cdots <s_{n-1}<\tau$. It follows from (\ref{eqnztau}) and the definitions of $z(\tau,n)$ and $F_n(\tau)$ that
$$
F_{n}(\tau) = \log \int\displaylimits_{0<s_1<\cdots <s_{n-1}<\tau} e^{B_1(0,s_1)+\cdots + B_n(s_{n-1},\tau)} ds_1\cdots ds_{n-1}.
$$

The path integral formula for $z(\tau,n)$ shows that it equals the partition function for a particular semi-discrete directed polymer in a random Brownian environment, first studied by O'Connell-Yor \cite{OY}. In this interpretation, $\log z(\tau,n)$ is the {\it quenched free energy} of the model. See the reviews \cite{CSY,ICReview} for some background on directed polymers.

\subsubsection{Fredholm determinant}

Since $q^{x_n(t)+n}$ converges (as $q\to 1$ and under appropriate scaling) to $z(\tau,n)$ the $e_q$-Laplace transform of $q^{x_n(t)+n}$ converges to the Laplace transform of $z(\tau,n)$. Thus, taking the $q\to 1$ limit of Theorem \ref{THMqTASEPFRED} yields the following result, first proved as \cite[Theorem 5.2.11]{BorCor} (see also \cite[Theorem 1.17]{BCF}). An alternative route to proving this theorem utilizes O'Connell's work \cite{OCon} on Whittaker measure in conjunction with an identity proved in \cite{BCR}.
\begin{theorem}\label{THMOYLaplace}
Consider the fundamental solution to the semi-discrete SHE. For any $n\geq 1$, $\tau\geq 0$ and $u\in \C$ with $\Real(u)>0$
$$
\EE\Big[e^{-u e^{\frac{3}{2}\tau} z(\tau,n)}\Big] = \det\big(I+K^{SD}\big)_{L^2(C_0)},
$$
where  $C_0$ is a small contour around 0 and
$$
K^{SD}(v,v') = \frac{1}{2\pi \I} \int_{1/2 - \I\infty}^{1/2+\I\infty} \Gamma(-s)\Gamma(1+s)\, \frac{u^s e^{v\tau s+ \frac{s^2 \tau}{2}}}{s+v-v'} \left(\frac{\Gamma(v-1)}{\Gamma(s+v-1)}\right)^n ds.
$$
\end{theorem}

It is natural to wonder whether this theorem could be proved directly in an analogous manner to the proof of Theorem \ref{THMqTASEPFRED}. It is possible to compute similar moment formulas for $\EE\big[z(\tau,n_1)\cdots z(\tau,n_k)\big]$ (see \cite[Proposition 5.2.9]{BorCor} or \cite[Section 6.2]{BCS}). A natural route to compute the Laplace transform of $z(\tau,n)$ would be to write
$$
\EE\left[e^{\zeta z(\tau,n)}\right]  = \EE\left[\,\sum_{k=0}^{\infty} z(\tau,n)^k \frac{\zeta^k}{k!}\right] = \sum_{k=0}^{\infty} \EE\left[z(\tau,n)^k\right] \frac{\zeta^k}{k!},
$$
and use the formulas for $\EE\left[z(\tau,n)^k\right]$. Unfortunately, the last equality above is not true (the first is true as it just amounts to the Taylor expansion of the exponential). It is not always possible to interchange expectations and infinite summations. The moments of $z(\tau,n)$ grow super-exponentially (as we have already seen from the discussion on intermittency). Therefore, the right-hand series is divergent for all $\zeta$ despite the fact that the left-hand side is necessarily convergent for $\zeta$ with negative real part. This issue of moment indeterminacy is alleviated by lifting up to the level of $q$-TASEP, where the algebra and analysis work hand-in-hand.

\subsubsection{KPZ class asymptotics}

Theorem \ref{THMOYLaplace} is amenable to asymptotic analysis as was performed in \cite[Theorem 5.2.13]{BorCor} and \cite[Theorem 1.3]{BCF} yielding:
\begin{theorem}\label{THMsdSHEasy}
For all $\nu>0$,
$$
\lim_{\tau\to \infty} \PP\left(\frac{\log z(\tau,\nu\tau) - \tau \tilde\gamma_1(\nu)}{d(\nu) \tau^{1/3}} \leq s\right) = F_{{\rm GUE}}(s)
$$
where $d(\nu) = \left(-\nu \Psi''(s(\nu))/2\right)^{1/3}$ with $s(\nu) = \arg\inf_{s>0} \big(s-\nu \Psi(s)\big)$.
\end{theorem}

Earlier, \cite{SeppValko} proved an upper bound on the variance of $\log z(\tau,\nu \tau)$ of order $\tau^{2/3}$, consistent with the $\tau^{1/3}$ scale of fluctuations. This theorem provides a matching lower bound as well as the exact limiting distribution.

\subsection{Example 4: ASEP}

\begin{figure}
\includegraphics[width=12cm]{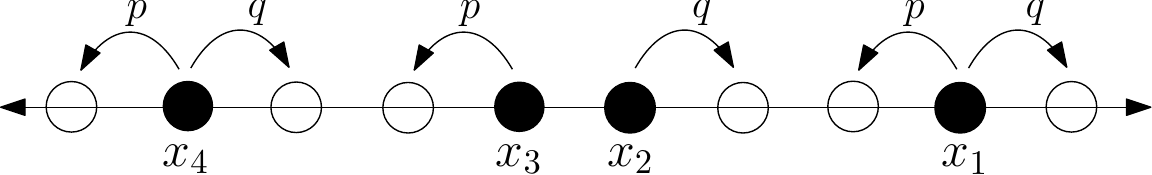}
\caption{ASEP particle configuration with possible jumps and rates denoted by arrows.}
\label{ASEP}
\end{figure}

The asymmetric simple exclusion process (ASEP) is a one-parameter deformation of TASEP in which particles can move both left and right. Let $x_i(t)\in \Z$ represent the location of particle $i$ at time $t$. We assume that $x_i(t)<x_j(t)$ for $i>j$. The state space of ASEP is the set of all such ordered $x_i$, and the dynamics can be described as follows: each particle $x_i$ has an exponential alarm clock (ringing after exponential waiting time, independent of all other particle clocks). When the alarm rings, the particle flips a coin and with probability $p$ attempts to jump left, and with probability $q=1-p$ attempts to jump right. The jump is achieved only if the destination site is unoccupied at that time. Regardless of the outcome, the particle's clock is immediately reset. We will assume that $0<p<q<1$ and $p+q=1$ so that there is a drift in the positive direction (like for TASEP in which $p=0$ and $q=1$). ASEP can also be interpreted in terms of a growing (and shrinking) height function: each $\vee$ is replaced by $\wedge$ at rate $q$ and each $\wedge$ is replaced by $\vee$ at rate $p$. As a measure of the asymmetry define the parameter  $\tau = p/q<1$ which will play a role akin to $q$ from $q$-TASEP. See Figure \ref{ASEP} for an illustration of ASEP.

Just as for TASEP and $q$-TASEP, we will work with step initial data in which $x_i(0)=-i$ for all $i\geq 1$ (and there are no other particles with lower labels).

\subsubsection{Moment formulas}\label{SECASEPmoments}

An observable of interest is the number of particles to have crossed a given site $y$. For $y\in \Z$, let $N_{y}(t) = \left|\left\{m\geq 1: x_m(t)\geq y\right\}\right|$. We would like to understand the behavior, in particular asymptotically, of this observable (which is closely related to the height in the growth interpretation of the model). Define $Q_y(t) = \tau^{N_y(t)}$ and its $\tau$-derivative
$$
\tilde{Q}_y(t) = \frac{Q_y(t)-Q_{y-1}(t)}{\tau-1}.
$$

\begin{theorem}\label{THMASEPmom}
Consider ASEP with step initial data and $\tau=p/q<1$. For any $k\geq 1$ and integers $y_1>\cdots >y_k$,
$$
\EE\left[\prod_{j=1}^{k} \tilde{Q}_{y_j}(t)\right] = \frac{\tau^{\frac{k(k-1)}{2}}}{(2\pi \I)^k} \oint\cdots \oint \prod_{1\leq A<B\leq k} \frac{z_A-z_B}{z_A-\tau z_B} \prod_{j=1}^{k} e^{-\frac{z_j(p-q)^2 t}{(1+z_j)(p+qz_j)}} \left(\frac{1+z_j/\tau}{1+z_j}\right)^{y_j+1} \frac{dz_j}{\tau +z_j}
$$
where the contours of integration are all along a small circle centered at $-\tau$ but not containing $-1$ or $-\tau^2$.
\end{theorem}

The limitation of having different $y_j$ in this theorem can be overcome by the identity
$$
Q_y^k(t) = \sum_{j=0}^{k} {n \choose x}_\tau (\tau;\tau)_j (-1)^j \sum_{y\leq y_1<\cdots < y_k} \prod_{j=1}^{k} \tilde{Q}_{y_j}(t).
$$
We have used the $q$-Pochhammer symbol and $q$-Binomial with $q$ replaced by $\tau$. This identity and Theorem \ref{THMASEPmom} enable us to recover a similar integral formula for $\EE\left[ \tau^{k N_y(t)}\right]$.
\subsubsection{Fredholm determinant}

Utilizing the methods described in Section \ref{SECFREDDETs} for $q$-TASEP, it is possible to turn the integral formula  for $\EE\left[ \tau^{k N_y(t)}\right]$ into a Fredholm determinant formula for the $e_\tau$-Laplace transform of $\tau^{N_y(t)}$ which first appeared as \cite[Theorem 5.3]{BCS}.

\begin{theorem}\label{THMASEPFREDDET}
Consider ASEP with step initial data and asymmetry parameter $\tau=p/q<1$. For any $y\in \Z$,
$$
\EE\left[\frac{1}{\big(\zeta \tau^{N_{y}(t)};\tau\big)_{\infty}}\right] = \det\big(I+K_{\zeta}^{ASEP}\big)_{L^2(C)}
$$
where
$$
K_{\zeta}^{ASEP}(w,w') = \frac{1}{2\pi \I} \int_D \Gamma(-s)\Gamma(1+s) (-\zeta)^s \frac{e^{(q-p)t \frac{\tau}{z+\tau}}\left(\frac{\tau}{z+\tau}\right)^{y}}{e^{(q-p)t \frac{\tau}{\tau^s z+\tau}}\left(\frac{\tau}{\tau^s z+\tau}\right)^{y}} \frac{-1}{q^s w-w'} \,ds.
$$
and the contours $C$ and $D$ can be found from the statement of \cite[Theorem 5.3]{BCS}.
\end{theorem}

\subsubsection{KPZ class asymptotics}

Theorem \ref{THMASEPFREDDET} characterizes the distribution of $N_y(t)$ and can be used to study its asymptotic behavior. There is another type of Fredholm determinant formula which can also be achieved from the moment formulas given earlier. That Fredholm determinant (known in \cite{BCS} as Cauchy-type) was essentially discovered earlier by Tracy-Widom \cite{TW1,TW2,TW3} using a different approach (a comparison of which is described in the short review \cite{twowaystosolve}). Asymptotic analysis performed in \cite{TW3} (and alternatively described in \cite[Section 9]{BCS}) yields:
\begin{theorem}
Consider ASEP with step initial data and asymmetry parameter $\tau=p/q<1$. Then
$$
\lim_{t\to \infty} \PP\left(\frac{N_0\big(t/(q-p)\big) - t/4}{t^{1/3}} \geq -r\right) = F_{{\rm GUE}}(2^{4/3}r).
$$
\end{theorem}

\subsection{Example 5: KPZ equation}\label{SECKPZ}

The Kardar-Parisi-Zhang (KPZ) equation was introduced in 1986 \cite{KPZ} by the eponymous trio of physicists as a continuous (in space and time) model of random interface growth. The height function $h:\R_{+}\times \R\to \R$ satisfies
$$
\frac{\partial}{\partial t} h(t,x) = \frac{1}{2} \frac{\partial^2}{\partial x^2} h(t,x) + \frac{1}{2} \left(\frac{\partial}{\partial x} h(t,x)\right)^2 + \xi(t,x)
$$
where $\xi(t,x)$ is space-time Gaussian white noise. In this continuous setting the Laplacian serves as a smoothing mechanism, the gradient squared serves as a mechanism for growth in the normal direction to the local slope, and the white noise inserts space-time uncorrelated randomness into the system. These three factors underly the KPZ universality class.

Making direct sense of this equation is challenging due to the non-linearity and the roughness of the spatial trajectories of $h$ (see \cite{BertiniCancrini, Hairer,Hairer2}). It has been understood since the work of \cite{BG} that the physically relevant notion of solution is to define
$$h(t,x):=\log z(t,x)$$
where $z:\R_{+}\times \R\to \R$ solves the well-posed stochastic heat equation (SHE) with multiplicative noise
$$
\frac{\partial}{\partial t} z(t,x) = \frac{1}{2} \frac{\partial^2}{\partial x^2} z(t,x) + \xi(t,x) z(t,x).
$$
The fundamental solution to the SHE has $z(0,x)=\delta_{x=0}$ and corresponds (under the weak scalings described in Section \ref{SECweak}) to step initial data.
(For more about this definition, see \cite{ICReview,ACQ}.) A variant of this {\it Hopf-Cole transform} between growth process and stochastic heat equation was already present in the context of the semi-discrete polymer in Section \ref{SEClimq}. Similar transforms also hold for $q$-TASEP and ASEP, amounting to the $k=1$ case of the dualities discussed later in Section \ref{SECqduality} and \ref{SECasepduality}, respectively.

The SHE has a directed polymer and parabolic Anderson model interpretation, though both require some care in making precise. Essentially, $z(t,x)$ can be interpreted as the partition function for a directed polymer model in which Brownian motion moves through a potential given by $\xi$ (see more in \cite{ACQ,AKQCDRP}) and can also be interpreted as the average mass density of a system of particles moving through $\R$ according to (independent) Brownian motions and splitting into two unit masses as well as dying according to the sign and amplitude of $\xi$.

\subsubsection{Weak scaling universality of the KPZ equation}\label{SECweak}
Rescale the solution to the KPZ equation by setting $h_{\e}(t,x) = \e^b h(\e^{-z}t, \e^{-1}x)$ where $b,z\in \R$. Then $h_{\e}$ satisfies
$$
\frac{\partial}{\partial t} h_{\e}(t,x) = \frac{1}{2} \e^{2-z}\frac{\partial^2}{\partial x^2} h_{\e}(t,x) + \frac{1}{2} \e^{2-z-b} \left(\frac{\partial}{\partial x} h_{\e}(t,x)\right) + \e^{b-z/2+1/2}\xi(t,x).
$$
Each term on the right-hand side rescaled differently.

Consider $b=1/2$ and $z=2$. Under this choice, the coefficients in front of the Laplacian and noise stay fixed as $\e$ varies, however the one in front of the squared gradient grows like $\e^{-1/2}$. If we inserted a parameter $\lambda$ in front of the squared gradient in the original KPZ equation, and simultaneously scaled $\lambda = \e^{1/2}$, then this would cancel the $\e^{-1/2}$ and the KPZ equation would remain invariant as $\e$ varied. We will call this {\it weak non-linearity scaling}.

Consider instead setting $b=0$ and $z=2$. Now, the coefficients in front of the Laplacian and squared gradient stay fixed as $\e$ varies, while the one in front of the white noise grows like $\e^{-1/2}$. Just as above, if we inserted a parameter $\beta$ in front of the white noise in the original KPZ equation, and scaled it as $\beta  = \e^{1/2}$, then this would cancel the $\e^{1/2}$ and the KPZ equation would remain invariant. We call this {\it weak noise scaling}.

These weak scalings are proxies for finding approximation schemes for the KPZ equation. Consider a model whose microscopic dynamics are characterized by a form of smoothing, a non-linear dependence of the growth rate on the local slope, and space-time uncorrelated noise. If either the non-linearity or the noise have tunable parameters, then applying the above weak scalings may yield convergence of the model to the KPZ equation. It is important to note that it is only under these special weak scalings that growth models are expected to converge to the KPZ equation. The KPZ universality class scaling demonstrated through the examples we have studied has $b=1/2$ and $z=3/2$, and does not involve a parameter scaling. One may be misled in taking a formal $\e\to 0$ limit of the rescaled KPZ equation with these choices of $b$ and $z$. It would seem that the (deterministic) inviscid Burgers equation arises as the limit, but this cannot be (for instance, we know the limit remains random). The non-linearity seems to enhance the noise, which formally disappears as $\e\to 0$. The KPZ-fixed point is the proposed \cite{CQ2} space-time limit of $h_{\e}(t,x)$ (and any KPZ class model under the same scaling). The $F_{{\rm GUE}}$ distribution is just a one-point marginal distribution for the fundamental solution to this fixed point evolution.

Returning to the weak scalings, $q$-TASEP, the semi-discrete SHE and ASEP all have tunable parameters which control either the strength of the non-linearity or the noise. They also all admit Hopf-Cole type transform to the form of SHEs (of course the semi-discrete SHE is already in such a form). Since the KPZ equation is defined via such a transform, this reduces the problem to proving convergence (under suitable weak scaling) of discrete SHEs to the continuous one. This was first achieved for ASEP in 1997 work of Bertini-Giacomin \cite{BG}, and subsequently has been extended to the other examples in \cite{ACQ,QMR} (and to discrete polymers in \cite{AKQ}). The only weak universality result which has not utilized an exact Hopf-Cole transform is that of \cite{Dembo} which deals with finite (jumps up to distance three) exclusion. That result still proceeds through a discrete SHE which is shown to closely approximate a Hopf-Cole type transformed height function.

\subsubsection{Moment formulas}
Limits of the moment formulas for $q$-TASEP, the semi-discrete random polymer and ASEP under weak scaling from Section \ref{SECweak} yield the following moment formula for the fundamental solution to the SHE.

\begin{theorem}\label{THMKPZMOMENTS}
Consider the fundamental solution to the SHE $z(t,x)$. For any $k\geq 1$ and $x_1\leq \cdots \leq x_k$
$$
\EE\left[\prod_{j=1}^{k} z(t,x_j)\right] = \frac{1}{(2\pi \I)^k} \int\cdots \int \prod_{1\leq A<B\leq k} \frac{z_A-z_B}{z_A-z_B-1} \prod_{j=1}^{k} e^{\frac{t}{2} z_j^2+x_j z_j} dz_j
$$
where the $z_j$ integration is over $\alpha_j + \I \R$ with $\alpha_1>\alpha_2 + 1>\alpha_3+ 2> \cdots$.
\end{theorem}
The moment Lyapunov exponents for $z(t,0)$ are easily computed from the above formula as $\gamma_k  = \frac{k^3-k}{24}$ (these were first computed by Kardar \cite{K} and proved in \cite{BertiniCancrini}).

\subsubsection{Fredholm determinant}\label{SecKPZFRED}

Just as for the semi-discrete random polymer, the moments of the SHE grow far to quickly to characterize the distribution of $z(t,x)$. However, we may use any of Theorems \ref{THMqTASEPFRED}, \ref{THMOYLaplace}, or \ref{THMASEPFREDDET} to prove the below Laplace transform formula for $z(t,x)$.

\begin{theorem}\label{THMKPZfred}
Consider the fundamental solution to the SHE. For any $\zeta\in \C$ with positive real part
$$
\EE\Big[e^{-\zeta e^{\frac{t}{24}} z(t,0)}\Big] = \det\big(I-K^{KPZ}_{\zeta}\big)_{L^2(\R_{+})},\qquad \textrm{where}\quad
K(\eta,\eta') = \int_{\R} \frac{\zeta}{\zeta + e^{-s(t/2)^{1/3}}} \Ai(s+\eta) \Ai(s+ \eta') ds.
$$
\end{theorem}

This Fredholm determinant can also be written in the same form as that of the earlier theorems. This formula (in fact the inversion of it giving the distribution of $z(t,0)$) was discovered independently and in parallel by Sasamoto-Spohn \cite{SaSp} and Amir-Corwin-Quastel \cite{ACQ} in 2010 based on asymptotic analysis of Tracy-Widom's ASEP formulas \cite{TW3}. The rigorous (mathematically) proof of the formula was provided by \cite{ACQ}, and another subsequent proof in \cite{BCF}. Soon after the work of \cite{SaSp,ACQ}, this formula was re-derived by Dotsenko \cite{Dot} and Calabrese-Le Doussal-Rosso \cite{CDR} via the mathematically non-rigorous replica method (i.e. using moments to try to recover the Laplace transform, despite the aforementioned impediments). For more details, consult \cite{ICReview}.

\subsubsection{KPZ class asymptotics}
A corollary of Theorem \ref{THMKPZfred} is that the KPZ equation is in the KPZ universality class. The below result was first proved in \cite[Corollary 1.3]{ACQ}. For stationary (i.e. $z(0,x) = B(x)$ a two-sided Brownian motion) initial data, the $t^{1/3}$ scale of fluctuations was demonstrated earlier in \cite{BQS}. Recently, using the KPZ line ensemble, \cite[Theorem 1.4]{CH2} show that this $t^{1/3}$ scale holds true for all KPZ initial data.

\begin{theorem}
Consider the fundamental solution to the SHE. For any $r\in \R$,
$$
\lim_{t\to \infty} \PP\left( \frac{\log z(t,0) + \frac{t}{24}}{(t/2)^{1/3}} \leq r\right) = F_{{\rm GUE}}(r).
$$
\end{theorem}

\subsection{Further examples}\label{SECfurtherexamples}

\begin{figure}
\includegraphics[width=14cm]{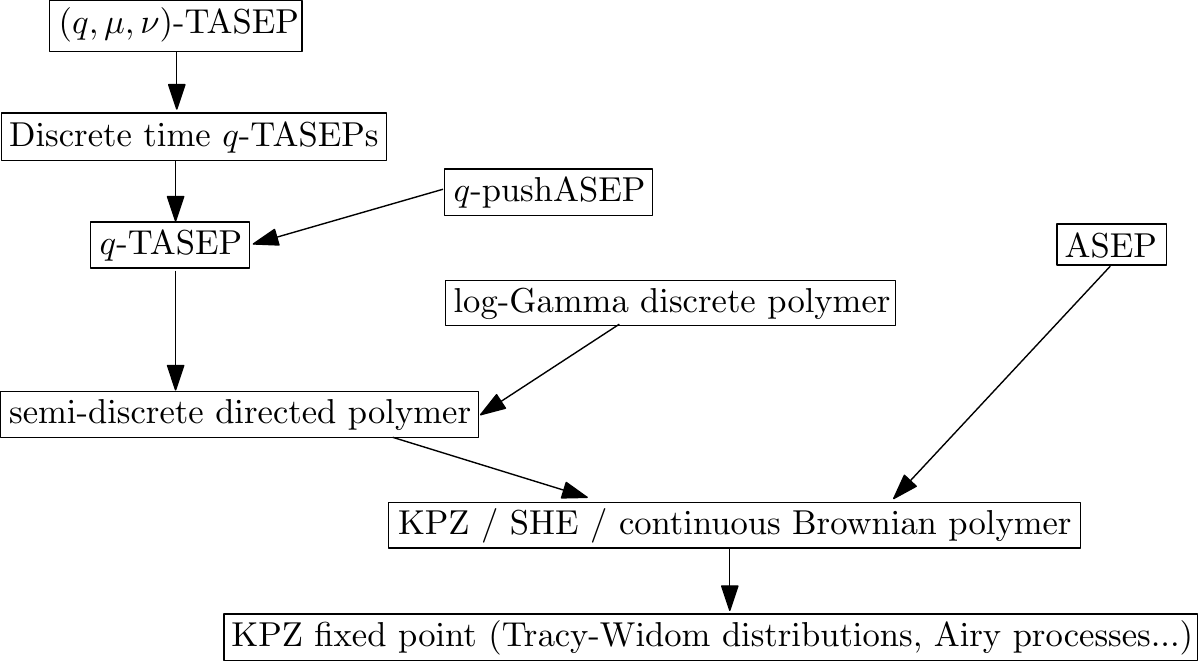}
\caption{Overview of (non-determinantal) integrable probabilistic systems (so far known) in the KPZ universality class.}\label{Processes}
\end{figure}

The list of (non-determinantal) integrable probabilistic systems in the KPZ universality class continues to grow. Figure \ref{Processes} records the names and relationships between these systems. In principal arrows should be transitive (though putting in the missing downward arrows requires either asymptotic analysis or stochastic analysis in each case). Besides those models we have already discussed in the examples, the $(q,\mu,\nu)$-TASEP has been studied in \cite{Cqmunu,Pov2}, the discrete time $q$-TASEPs in \cite{BCdiscrete}, the $q$-PushASEP in \cite{BorPet,CorPet}, and the log-gamma polymer in \cite{SeppLog,COSZ,OSZ}. It seems likely that the methods we now turn to will yield the discovery and analysis of further examples beyond these.

\section{Macdonald processes}\label{SECMAC}

A high point of modern representation theory and symmetric function theory, Macdonald symmetric polynomials have found many diverse applications throughout mathematics. The canonical reference for their properties is the book \cite{M} (see also the review material in \cite[Section 2]{BorCor}, and the historical perspective at the end of \cite{BorProc}). In this section we present a probabilistic application of these remarkable polynomials.

\subsection{Defining Macdonald symmetric polynomials}\label{SECdefmac}
Macdonald symmetric polynomials in $N$ variables $x_1,\ldots, x_N$ are indexed by non-negative integer partitions $\lambda = (\lambda_1\geq \cdots \geq \lambda_N\geq 0)$ and written as $P_{\lambda}(x_1,\ldots, x_N)$. They are invariant under the action of the symmetric group $S_N$ on the $N$ variables, and have coefficients which are rational functions of two additional parameters $q,t$ (i.e. coefficients in $\Q(q,t)$) which we assume are in $[0,1)$. The $P_{\lambda}$ (as $\lambda$ varies) form a linear basis in symmetric polynomials in $N$ variables over $\Q(q,t)$. They can be defined in the following (rather inexplicit) manner (which will, however, suffice for our purposes). Define the {\it Macdonald first difference operator} $D^N_1$ on the space of $N$ variable symmetric functions $f$ as
$$
\big(D^N_{1} f\big)(x_1,\ldots, x_N) = \sum_{i=1}^{N} \prod_{\stackrel{j=1}{j\neq i}}^{N} \frac{t x_i-x_j}{x_i-x_j} f(x_1,\ldots ,qx_i, \ldots, x_N).
$$
It is not a priori clear (due to the denominator $x_i-x_j$), but this operator preserves the class of symmetric polynomials. This operator is self-adjoint (with respect to a natural inner product on symmetric polynomials with coefficients in $\Q(q,t)$) and the Macdonald symmetric polynomials are the eigenfunctions of $D^N_1$ labeled via their (generically) pairwise different eigenvalues
$$
\big(D^N_1 P_{\lambda}\big)(x_1,\ldots, x_N)  = (q^{\lambda_1} t^{N-1} + q^{\lambda_2} t^{N-2} +\cdots + q^{\lambda_N})P_{\lambda}(x_1,\ldots, x_N).
$$
The polynomials have many striking properties. They are orthogonal (as eigenfunctions of $D^N_1$) with respect to the earlier mentioned inner product, and the Macdonald $Q_{\lambda}$ polynomials are defined as $P_{\lambda} / \langle P_{\lambda},P_{\lambda}\rangle$ and form a dual basis to the $P_{\lambda}$. There is a Cauchy type identity providing a simple reproducing kernel: for variables $a_1,\ldots, a_N$ and $b_1,\ldots, b_M$ with $|a_ib_j|<1$ for all $i,j$,
$$
\sum_{\lambda}P_{\lambda}(a_1,\ldots, a_N) Q_{\lambda}(b_1,\ldots, b_M) = \prod_{i,j} \frac{(t a_ib_j;q)_{\infty}}{(a_ib_j;q)_{\infty}} =: \Pi(a_1,\ldots, a_N;b_1,\ldots, b_M).
$$
They satisfy Pieri and branching rules: the first describes the coefficients which result from multiplying Macdonald symmetric polynomials by elementary (or $(q,t)$-complete homogeneous) symmetric polynomials and reexpressing the answer in terms of other Macdonald symmetric polynomials; the second will be described below in Section \ref{SECconstdyn}. In the results explained below, these are essentially the only properties of these polynomials utilized. Other noteworthy properties are index/variable duality, and the existence of $N-1$ other difference operators which commute with $D^N_1$ (and also are diagonalized by the $P_{\lambda}$).

\subsection{Defining Macdonald processes}

\begin{figure}
\includegraphics[height=5cm]{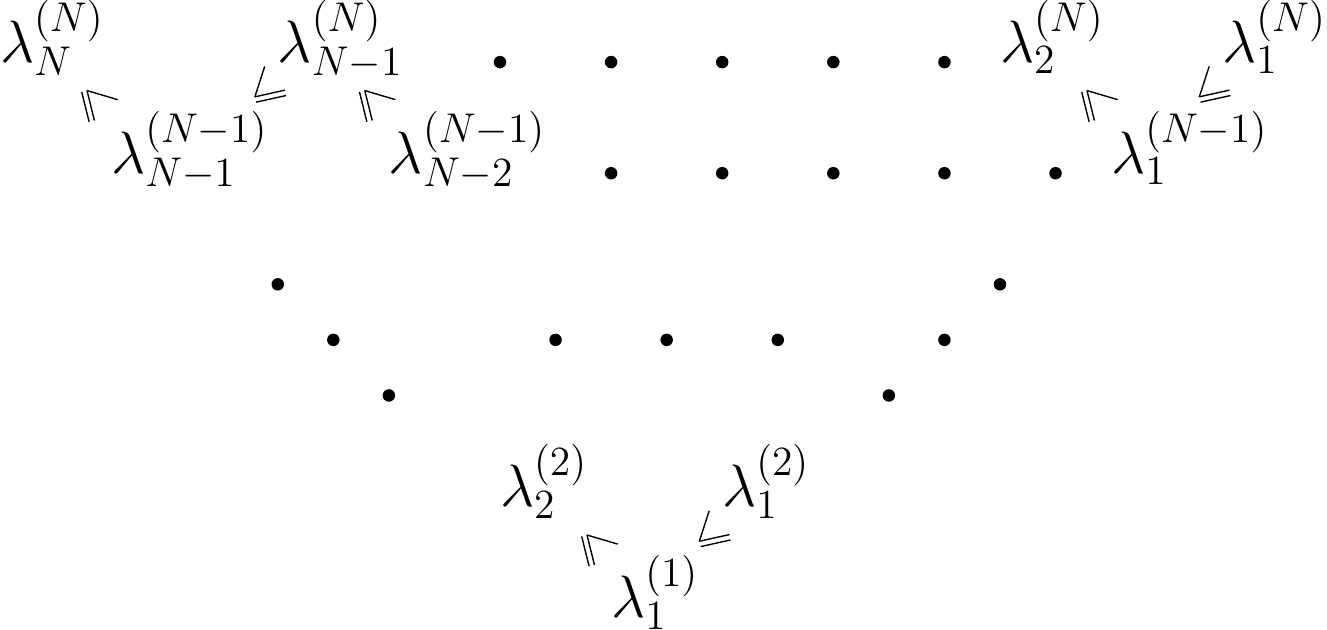}
\caption{A Gelfand-Tsetlin pattern of interlacing non-negative integer partitions.}
\label{GTPattern}
\end{figure}

The {\it (ascending) Macdonald process} is a probability measures on interlacing partitions $\lambda^{(N)}\succeq\lambda^{(N-1)}\succeq\cdots \succeq\lambda^{(1)}$ where the number of non-zero elements in $\lambda^{(m)}$ is at most $m$, and the symbol $\succeq$ implies interlacing (so $\lambda^{(m)}_{j+1} \leq \lambda^{(m)}_{j} \leq \lambda^{(m+1)}_{j}$ for all meaningful inequalities). Such an interlacing triangular arrays of non-negative integers is also known as a {\it Gelfand-Tsetlin pattern}. See Figure \ref{GTPattern} for such an array.

Measures on interlacing triangular arrays arise in many contexts. Before defining Macdonald processes, we consider a simpler example which comes from random matrix theory. Consider an $N\times N$ Gaussian Hermitian matrix  drawn from the Gaussian unitary ensemble. For any $m\leq N$ let $\lambda^{(m)}_{1},\ldots, \lambda^{(m)}_{m}$ be the ordered (largest to smallest) eigenvalues of the $m\times m$ upper-left corner of the matrix. By Rayleigh's Theorem (see, for instance \cite{BH}) the eigenvalues at level $m$ interlace with those at level $m-1$. Thus, the eigenvalues form an interlacing triangular array, though the constitute elements are reals now instead of non-negative integers. The measure on this array inherited from the GUE measure is called the GUE-corner (or sometimes minor) process \cite{Bar,Nord} and has a very nice form. At level $N$, the measure on the eigenvalues $\lambda^{(N)}$ is the GUE measure written as (up to normalizations)
$$\prod_{i\neq j}^{N} (\lambda^{(N)}_i - \lambda^{(N)}_j)^2 \prod_{i=1}^{N} e^{-(\lambda^{(N)}_i)^2/2}.$$
Given the eigenvalues at level $N$, the distribution of $\lambda^{(N-1)},\ldots, \lambda^{(1)}$ is uniform over the Euclidean simplex such that the interlacing inequalities are all satisfied \cite{GN,Bar,Ne}.

The Macdonald process is a far reaching generalization of the GUE-corner process. In order to describe it we will start by describing the Macdonald analog of the GUE measure on level $N$. This single level measure is called the Macdonald measure and defined as
$$
\MM_{(N;a,b)}\big(\lambda^{(N)}\big) := \frac{P_{\lambda^{(N)}}(a_1,\ldots, a_N) Q_{\lambda^{(N)}}(b_1,\ldots, b_M)}{\Pi(a_1,\ldots, a_N;b_1,\ldots, b_M)}.
$$
Here $a=(a_1,\ldots, a_N)$ and $b=(b_1,\ldots, b_M)$ for some $M\geq 0$ (one can work with more general Macdonald non-negative specializations -- see \cite{BorCor,BCGS}). From the Cauchy type identity, it is clear that summing over all $\lambda^{(N)}$ yields one. If the $a_i$ and $b_j$ are all non-negative, then, due to a combinatorial expansion formula for the $P_{\lambda}$ and $Q_{\lambda}$, the numerator (and thus the measure) is also non-negative. Besides the dependence on the $a$ and $b$ parameters, the measure also depends on the Macdonald $q,t$ parameters. We will hold off defining the Macdonald process until Section \ref{SECconstdyn}.

As shown in Figure \ref{Picture}, Macdonald process generalizes a number of other measures. The GUE measure / GUE-corner process is a continuous space degeneration of the Schur measure / process \cite{OkSchur,OkResh,Nord,OkResh2}. Macdonald measure seems to have first studied by Fulman in 1997 \cite{Fulman}, and subsequently by \cite{FR05,Vuletic09}. Until recently there were few examples of interesting probabilistic systems related to the Macdonald measure / process and there was a lack of ways to compute with them. In short, we were generally missing the answers to the questions of why and how to study Macdonald processes.

In 2011, Borodin-Corwin \cite{BorCor} provided partial answers to these two questions by:
\begin{enumerate}
\item constructing explicit Markov operators that map Macdonald processes to Macdonald processes (with updated parameters);
\item evaluating averages of a rich class of observables of the measures.
\end{enumerate}
In both cases, the integrable structure of Macdonald polynomials translates directly into probabilistic content.

Since the work of \cite{BorCor}, there has been a flurry of activity in these directions (see Section \ref{SECFURTHMAC}). We will only touch on the simplest example of how both of these answers work.

\subsection{Computing expectations}\label{SECcomputeexpect}
Within statistical mechanics it is desirable to find explicit formulas for ensemble partition functions. For example, for the Ising model (at inverse temperature $\beta$ in magnetic field $h$) the partition function is $Z(\beta,h) = \sum_{\sigma} e^{\beta \sum_{i\sim j} \sigma_i \sigma_j + h\sum_i \sigma_i}$. Taking derivatives of $\log Z(\beta,h)$ in $h$ and $\beta$ give (respectively) the expected magnetization, and expected product of spin over neighboring sites. The key here is that the Boltzmann weight (inside the sum over spin configurations $\sigma$) is an eigenfunction for the operators of differentiation in $h$ and in $\beta$.

In our present case $\Pi(a;b)$ is like the partition function and $P_{\lambda}(a)Q_{\lambda}(b)$ the Boltzmann weight. Let $D$ be any linear operator which is diagonalized by the Macdonald polynomials (e.g. a product of the Macdonald difference operators) with eigenvalue $d_{\lambda}$, so that.
$$
\big(DP_{\lambda}\big)(a) = d_{\lambda} P_{\lambda}(a).
$$
Since $\sum_{\lambda} P_{\lambda}(a) Q_{\lambda}(b) = \Pi(a;b)$, it follows that (with $D^{(a)}$ meaning to apply $D$ on the $a$ variables)
$$
D^{(a)}\Pi(a;b) = \sum_{\lambda} D^{(a)}P_{\lambda}(a)Q_{\lambda}(b) = \sum_{\lambda} d_{\lambda} P_{\lambda}(a)Q_{\lambda}(b).
$$
Dividing both sides by $\Pi(a;b)$ yields
$$
\EE\big[d_{\lambda}\big]  = \frac{D^{(a)}\Pi(a;b)}{\Pi(a;b)}.
$$
If all of the ingredients are explicit (as they are for products of Macdonald difference operators), then we obtain meaningful and explicit probabilistic information without ever needing to know explicit formulas for the Macdonald measure itself. In fact, the eigenvalues of the commuting family of Macdonald difference operators provide explicit formulas for expectations of enough observables to entirely characterize the Macdonald measure. In this way, the Macdonald measure is a completely integrable probabilistic system.

We will return to one such explicit formula (with Macdonald parameter $t=0$) in Section \ref{SECexpexample}, and refer readers to \cite[Section 2.2.3]{BorCor} and \cite{BCGS} for a more general discussion of developments here.

\subsection{Constructing dynamics}\label{SECconstdyn}

The construction of dynamics on Gelfand-Tsetlin patterns which we present comes from an idea of Diaconis-Fill \cite{DiaconisFill} in 1990 and was developed in the case of Schur processes by Borodin-Ferrari \cite{BF} in 2008 (see all \cite{twosides}). Before describing this construction we explain how the full Macdonald process is defined (we have so far only defined the Macdonald measure on a given level $N$).

The branching rule for $P_{\lambda^{(N)}}$ is
$$
P_{\lambda^{(N)}}(a_1,\ldots, a_{N-1},a_N) = \sum_{\lambda^{(N-1)} \preceq \lambda^{(N)}} P_{\lambda^{(N-1)}}(a_1,\ldots, a_{N-1}) P_{\lambda^{(N)}/\lambda^{(N-1)}}(a_N)
$$
where the sum is over all partitions $\lambda^{(N-1)}$ which interlace with $\lambda^{(N)}$ and where the skew Macdonald polynomial $P_{\lambda/\mu}(u)$ is zero unless $\lambda \preceq \mu$ and $\psi_{\lambda/\mu} u^{|\lambda|-|\mu|}$ otherwise (with $\psi_{\lambda/\mu}\in \Q(q,t)$ explicit and not dependent on $u$).

It follows from the branching rule that the Markov kernel (or stochastic link) $\Lambda^{N}_{N-1}$ from level $N$ to level $N-1$ given by
$$
\Lambda_{N-1}^{N}\big(\lambda^{(N)},\lambda^{(N-1)}\big) := \frac{P_{\lambda^{(N-1)}}(a_1,\ldots, a_{N-1}) P_{\lambda^{(N)}/\lambda^{(N-1)}}(a_N)}{P_{\lambda^{(N)}}(a_1,\ldots, a_N)}
$$
maps the Macdonald measure $\MM_{(N;a_1,\ldots, a_{N-1},a_N;b)}$ on level $N$ to the Macdonald measure $\MM_{(N;a_1,\ldots, a_{N-1};b)}$ on level $N-1$ (note that the $a_N$ has been removed). The law of the trajectory of the (inhomogeneous) Markov chain defined by these kernels and started from Macdonald measure on level $N$ is the Macdonald process. In other words, the Macdonald process on $\lambda^{(N)}\succeq \cdots \succeq \lambda^{(1)}$ specified by parameters $a_1,\ldots, a_N$ and $b_1,\ldots, b_M$ is written as $\MM_{([1,N];a;b)}$ and defined as
$$
\MM_{([1,N];a;b)}\big(\lambda^{(N)},\ldots,\lambda^{(1)}\big) := \MM_{(N;a;b)}\big(\lambda^{(N)}\big) \Lambda^{N}_{N-1}\big(\lambda^{(N)},\lambda^{(N-1)}\big) \cdots \Lambda^{(2)}_{1}\big(\lambda^{(2)},\lambda^{(1)}\big).
$$

In the GUE-corner process, the stochastic link $\Lambda^N_{N-1}$ is given by the indicator function that $\lambda^{(N-1)}$ interlaces with $\lambda^{(N)}$ times the ratio of the volume of the simplex of triangular arrays with top level $\lambda^{(N-1)}$ to that with top level $\lambda^{(N)}$ (there volumes are given by Vandermonde determinants).

There is another natural Markov chain which maps Macdonald measure to Macdonald measure on a single level $N$. For $u\geq 0$, the Markov kernel
$$
\pi^{(u)}_{N}\big(\lambda^{(N)},\nu^{(N)}\big) := \frac{P_{\nu^{(N)}}(a)}{P_{\lambda^{(N)}}(a)} \cdot \frac{Q_{\nu^{(N)}/\lambda^{(N)}}(u)}{\Pi(a;u)}
$$
maps the Macdonald measure $\MM_{(N;a;b_1,\ldots, b_M)}$ on level $N$ to the Macdonald measure $\MM_{(N;a;b_1,\ldots, b_M,u)}$ on level $N$ (note that $u$ has been appended to the $b$-list). This Markov kernel has the interpretation as the Doob-h transform of the sub-Markov kernel given by $\frac{Q_{\nu^{(N)}/\lambda^{(N)}}(u)}{\Pi(a;u)}$. Due to the explicit formula for the skew Macdonald polynomial, this sub-Markov kernel acts on $\lambda^{(N)}$ by increasing each element by independent geometrically distributed (with parameter $u$) amounts, and then killing all configurations which violate interlacing, and energetically penalizing all other configurations based on the value of $Q_{\nu^{(N)}/\lambda^{(N)}}(u)$. A generalized Cauchy type identity implies
$$
\sum_{\nu^{(N)}} \frac{Q_{\nu^{(N)}/\lambda^{(N)}}(u)}{\Pi(a;u)} P_{\nu^{(N)}}(a) = P_{\lambda^{(N)}}(a)
$$
hence $P_{\nu^{(N)}}(a)$ has eigenvalue one for this sub-Markov kernel and is positive inside and zero outside the support of the kernel. The Markov kernel $\pi^{(u)}_{N}$ therefore corresponds to conditioning the sub-Markov chain to survive forever.

In the GUE setting, and in continuous time, this Markov chain is replaced by Dyson's Brownian motion (which can be thought of as conditioning $N$ Brownian motions to never intersect). Therefore, the Markov chain corresponding to $\pi^{(u)}_{N}$ is a discrete time $(q,t)$-deformation of Dyson's Brownian motion.

We have defined two Markov chains. One chain goes from level $N$ to level $N-1$ with kernel $\Lambda_{N-1}^{N}$ and the other goes from level $N$ to level $N$ with kernel $\pi^{(u)}_{N}$. We will introduce a multivariate Markov chain with state space given by the entire Gelfand-Tsetlin pattern that `stitches' these two chains together.

\begin{figure}
\includegraphics[height=7cm]{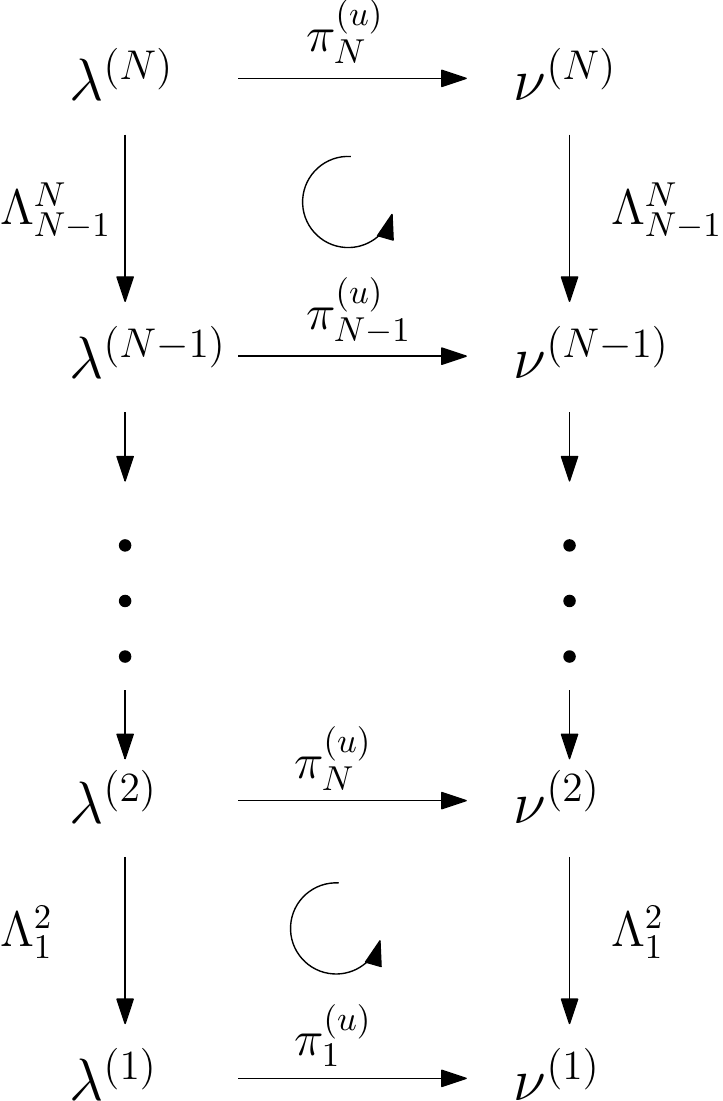}
\caption{Intertwining Markov kernels allow for construction of multivariate dynamics preserving Macdonald measure.}
\label{Commute}
\end{figure}

The key input into this construction is an intertwining relation of the two Markov chains. Specifically, for $u\geq 0$, and any $m\geq 2$, $\Lambda^{m}_{m-1} \pi^{(u)}_{m-1} = \pi^{(u)}_{m}\Lambda^{m}_{m-1}$. This intertwining is illustrated in Figure \ref{Commute}.

For $u\geq 0$ define the Markov kernel
$$
P^{(u)}\big((\lambda^{(1)},\ldots, \lambda^{(N)}),(\nu^{(1)},\ldots, \nu^{(N)})\big) := \pi^{(u)}_1\big(\lambda^{(1)},\nu^{(1)}\big) \prod_{k=2}^{N} \frac{\pi^{(u)}_{k}\big(\lambda^{(k)},\nu^{(k)}\big) \Lambda^{k}_{k-1}\big(\nu^{(k)},\nu^{(k-1)}\big)}{\big(\pi^{(u)}_{k}\Lambda^{k}_{k-1}\big) \big(\lambda^{(k)},\nu^{(k-1)}\big)}.
$$
Then $P^{(u)}$ maps the Macdonald process $\MM_{([1,N];a;b_1,\ldots, b_M)}$ to the Macdonald process $\MM_{([1,N];a;b_1,\ldots, b_M,u)}$. The important property of this construction is that each level $m$ marginally evolves according to $\pi^{(u)}_{m}$, while the entire chain preserves the structure of the Macdonald process. $P^{(u)}$ first updates $\lambda^{(1)}$ to $\nu^{(1)}$ based on $\pi^{(u)}_1$, then updates $\lambda^{(2)}$ to $\nu^{(2)}$ according to the conditional law of $\nu^{(2)}$ given that the $\Lambda^{2}_1$ transition should bring $\nu^{(2)}$ to the previously determined $\nu^{(1)}$. The update proceeds similarly on each sequential pair of levels. These dynamics are constructed in \cite[Section 2.3]{BorCor} and further constructions of dynamics which preserve the  class of Macdonald processes (or their degenerations) are given in \cite{BorPet,OConPei,OCon,COSZ}.

In the GUE setting, and in continuous time, the limit (cf. \cite{GorinShkol,GorinShkol2}) of this construction yields Warren's process \cite{Warren} in which $\lambda^{(1)}_1$ evolves as a Brownian motion, $\lambda^{(2)}_1$ and $\lambda^{(2)}_2$ evolve according to independent Brownian motions which are reflected above and below (respectively) $\lambda^{(1)}_1$, and in general $\lambda^{(m)}_j$ evolves as a Brownian motion reflect to be above $\lambda^{(m-1)}_j$ and below $\lambda^{(m-1)}_{j-1}$. These dynamics preserve the class of GUE corner processes and have GUE Dyson's Brownian motion marginally on each level.

\subsection{Example of dynamics}

The dynamics constructed in Section \ref{SECconstdyn} becomes simpler when we set the Macdonald parameter $t=0$ and move into a continuous time setting. Since from here on out the Macdonald parameter $t$ is fixed to be zero, we will abuse notation and use $t$ for time. This transition to continuous time is achieved through setting the parameter $u$ in the construction equal to $\e(1-q)$ and running the discrete time Markov dynamics for $\e^{-1} t$ steps (the factor of $1-q$ makes formulas nicer). Taking $\e\to 0$ yields the following continuous time (measure by $t$) dynamics.

Treat the $\lambda^{(m)}_{k}$ as coordinates of particles where $m$ is the level on which they live and $k$ is their horizontal location. Each particle $\lambda^{(m)}_{k}$ jumps by one horizontally to the right independent of the others according to an exponential clock of rate
\begin{equation}\label{eqnrates}
\textrm{rate}(\lambda^{(m)}_{k}) = a_m \frac{\left(1-q^{\lambda^{(m-1)}_{k-1} - \lambda^{(m)}_{k}}\right)\, \left(1-q^{\lambda^{(m)}_{k} - \lambda^{(m)}_{k+1}+1}\right)}{\left(1-q^{\lambda^{(m)}_{k} - \lambda^{(m-1)}_{k}+1}\right)}.
\end{equation}
Those of the three terms above which refer to particles labeled with $m=0$, $k=m+1$ or $k=0$ are simply left out of the formula.

\begin{figure}
\includegraphics[height=3cm]{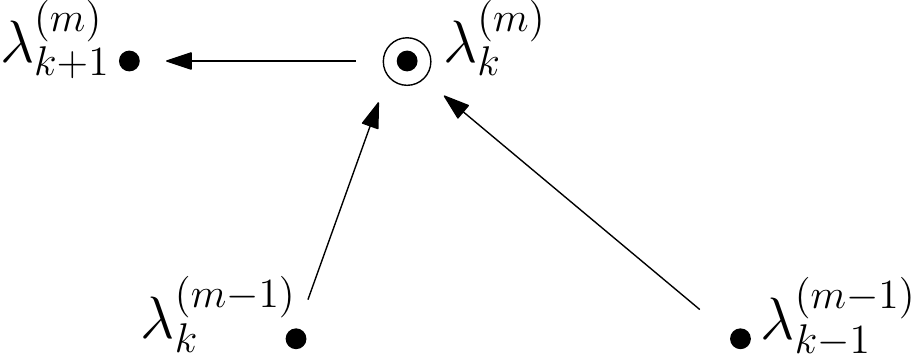}
\caption{The interactions felt by $\lambda^{(m)}_k$.}
\label{Rates}
\end{figure}

This is a $(2+1)$-dimensional interacting particle system with a local (in terms of particle labels) update rule. The particle $\lambda^{(m)}_{k}$ is influenced by the horizontal distance to three of its neighbors. As it gets closer to $\lambda^{(m-1)}_{k-1}$, its jump rate slows to zero (preventing jumps out of the interlacing condition). As it gets closer to $\lambda^{(m-1)}_{k}$ the jump rate increases to infinity (so as to immediately force a jump if $\lambda^{(m-1)}_{k}$ has overtaken the particle). These two interactions are the strongest, however there is also a slow down as $\lambda^{(m)}_{k+1}$ gets closer to the particle. These forces are illustrated in Figure \ref{Rates}.

By virtue of (\ref{eqnrates}), the set of coordinates $\big\{\lambda^{(m)}_{m} - m\big\}_{m\geq 1}$ evolves autonomously of the rest of the Gelfand-Tsetlin pattern. This $(1+1)$-dimensional interacting particle system is $q$-TASEP where $x_m(t) = \lambda^{(m)}_{m}$ at time $t$ and where the jump rate of $x_{m}$ is given by $a_m(1-q^{x_{m-1}(t)-x_{m}(\tau)-1})$. We have been led to this particle system by virtue of the properties of Macdonald polynomials.

Step initial data for $q$-TASEP is achieved by running the above dynamics on Gelfand-Tsetlin patterns from initial data given by $\MM_{([1,N];a;0)}$. By setting all $b_j$ to be zero, this measure is entirely supported on the configuration where all $\lambda^{(m)}_{k} \equiv  0$. After performing the above affine shift to $x_m$ coordinates, this corresponds with setting $x_m(0)=-m$ for $m\geq 1$.

\subsection{Example of expectations}\label{SECexpexample}

Running the continuous time (Macdonald parameter $t=0$) dynamics for time $t$ (recall our abuse of notation) from initial data given $\MM_{([1,N];a;0)}$ yields another Macdonald process, which can be thought of as the $\e\to 0$ limit of $\MM_{([1,N];a;\e(1-q),\cdots \e(1-q))}$ where there are $\e^{-1}t$ entries of $\e(1-q)$. This limit is called the {\it Plancherel specialization} and denoted by $\rho_{t}$ so that the limiting measure becomes $\MM_{([1,N];a;\rho_{t})}$. Under this limit $\Pi(a;b)$ becomes
$$
\Pi(a;\rho_{t}) = \prod_{i=1}^{N} e^{a_i t}.
$$

We will now utilize the prescription of Section \ref{SECcomputeexpect} to compute observables of this Macdonald process (and hence also of $q$-TASEP started from step initial data). As we have fixed the Macdonald parameter $t=0$, the eigenvalue of the first difference operator simplifies so that $D^N_1 P_{\lambda^{(N)}}(a) = q^{\lambda^{(N)}_N} P_{\lambda^{(N)}}(a)$. Therefore,
$$
\EE\left[q^{k (x_N(t)+N)}\right] = \EE\left[q^{k\lambda^{(N)}_N}\right] = \frac{(D^N_1)^k \Pi(a;\rho_{t})}{\Pi(a;\rho_{t})}
$$
where the first expectation is over $q$-TASEP started from step initial data and the second expectation is over the Macdonald process $\MM_{([1,N];a;\rho_{t})}$. This can be generalized \cite{BCGS} to any $n_1\geq \cdots \geq n_k\geq 1$
$$
\EE\left[\prod_{j=1}^{k} q^{x_{n_j}(\tau)+n_j}\right] = \EE\left[\prod_{j=1}^{k} q^{\lambda^{(n_j)}_{n_j}}\right] = \frac{D^{n_k}_1\cdots D^{n_1}_1 \Pi(a;\rho_{t})}{\Pi(a;\rho_{t})}
$$
where $D^{n}_1$ represents the Macdonald first difference operator applied only to the variables $a_1,\ldots, a_n$. Theorem \ref{THMqTASEPmoments} follows (in fact a general $a_i$ version of it) via encoding the application of these difference operators in terms of residues from contour integrals. To state the general $a_i$ formula, assume (for simplicity of the choice of contours) that all $a_i$ are very close to 1. Then using the multiplicative form of $\Pi(a;\rho_t)$ we find that
$$
\frac{D^{n_k}_1\cdots D^{n_1}_1 \Pi(a;\rho_{t})}{\Pi(a;\rho_{t})} = \frac{(-1)^k q^{\frac{k(k-1)}{2}}}{(2\pi \I)^k}\oint \cdots \oint \prod_{1\leq A<B\leq k} \frac{z_A-z_B}{z_A-q z_B} \prod_{j=1}^{k} \prod_{m=1}^{n_j} \frac{a_m}{a_m-z_j} e^{(q-1) t z_j} \frac{dz_j}{z_j}
$$
where, for each $A\in \{1,\ldots,k\}$ the contour of integration of $z_A$ contains the set of all $a_i$, as well as $q$ times the contour of integration of $z_B$ for $B>A$, but does not contain $0$. Computing residues as the $z_k$ through $z_1$ contours are shrunk provides a direct link to the difference operators.

\subsection{Further developments}\label{SECFURTHMAC}

\begin{figure}
\centering\includegraphics[trim=0cm 1cm 0cm 0cm, clip=true, width=16cm]{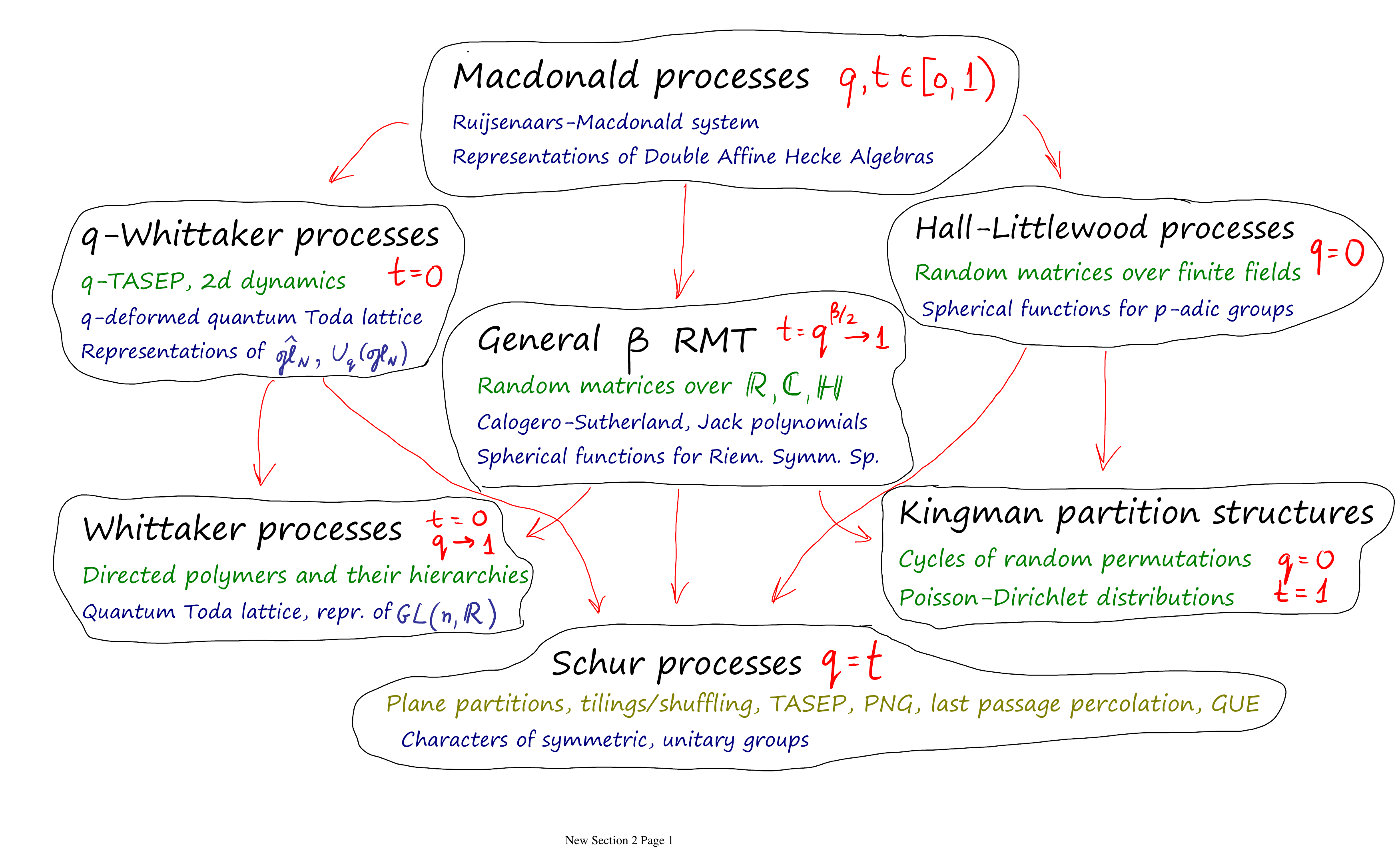}
\caption{Macdonald processes and their degenerations}\label{Picture}
\end{figure}
We record (without description) some further developments related to the theory of Macdonald processes. Figure \ref{Picture} highlights and organizes some of the probabilistic systems related to Macdonald processes and their degenerations (limits under special choices of Macdonald $q,t$ parameters). These degenerations mimics those of Macdonald symmetric polynomials. The Schur process degeneration has been well studied during the past decade (see the review \cite{BorGorreview} or \cite{BorProc}) so we forego further discussion below. Note, in \cite{Amol} the Macdonald process technology is utilized to rederive the Schur process determinantal structure. We also forego discussion of Kingman partition structures and refer the interested reader to \cite{Petrov} and references therein.

\begin{itemize}
\item Further dynamics have been constructed which preserve Macdonald processes (or their degenerations) \cite{BorCor, OCon, COSZ, BorPet, OConPei, BorGorBeta, GorinShkol2, BufPet}.
\item Further probabilistic systems have been connected to Macdonald processes (or their degenerations) or discovered via the above dynamics \cite{BorCor, OCon, COSZ,BorPet, OSZ, CorPet, BCdiscrete, BufPet, BorGorBeta}.
\item Exact and concise formulas have been found for expectations for a rich class of observables \cite{BorCor, BCGS, OCon, COSZ, BCR, BCF, BCFV, BorGorBeta}.
\item A formal power series treatment of Macdonald processes and observable formulas has been developed \cite{BCGS}.
\item Asymptotic analysis has been performed on some of the systems related to Macdonald processes \cite{BorCor, BCF, BCFV, BCR, FerVet, BCLyapunov, BorGorBeta}.
\item Some of the structure related to Macdonald processes has been recast in the probabilistic language of Gibbsian line ensembles and used to prove some universality results beyond exact solvable situations \cite{CH2}.
\item Formulas for expectations as well as dynamics preserving Macdonald process have begun to be connected to the Bethe ansatz and theory of quantum integrable systems \cite{BCS,BCdiscrete, CorPet}.
\end{itemize}

It is this last point, the connection to Bethe ansatz and quantum integrable systems, which we expand upon in Section \ref{SECQIS}.

\section{Quantum integrable systems}\label{SECQIS}

We will not define a quantum integrable system or go into any depth as to their algebraic origins (see \cite{Fadeev, QISM, Baxter, Resh} for some references in this direction). Instead, we will study a few systems which arise in relation to the probabilistic analysis of models in the KPZ universality class.

\subsection{Delta Bose gas}
The first connection between the KPZ universality class and a quantum integrable system came from independent work of Kardar \cite{K} and Molchanov \cite{Mol} in 1987. For the SHE $z(t,x)$ (recall from Section \ref{SECKPZ}) joint moments $\EE\big[z(t,x_1)\cdots z(t,x_k)\big]$ are solutions to the quantum delta Bose gas, or Lieb-Liniger model (in imaginary time and with attractive delta interaction):
$$
\frac{\partial}{\partial t} \EE\big[z(t,x_1)\cdots z(t,x_k)\big] = \frac{1}{2} \left(\sum_{i=1}^{k} \frac{\partial^2}{\partial x_i^2} + \prod_{i\neq j}^{k} \delta(x_i-x_j)\right) \EE\big[z(t,x_1)\cdots z(t,x_k)\big].
$$
In 1963 Lieb-Liniger solved (i.e. computed eigenfunctions for) the Hamiltonian on the right-hand side (i.e. the operator in the parentheses) via the Bethe ansatz (see also \cite{McGuire,Yang1,Yang2} expanding on this initial work). This was the second instance of a model being solved via this method, the first being Bethe's original solutions to the spin $1/2$ $XXX$ Heisenberg chain. Lieb-Liniger's work marked the beginning of the development of the theory of quantum integrable systems. Besides computing eigenfunctions, for many purposes it is necessary to prove the completeness of the Bethe ansatz and determine the norms of the eigenfunctions. Such results go under the general title of {\it Plancherel theorems} and we will return to discuss these as well as the Bethe ansatz in Section \ref{SECcoord}.

Using the eigenfunctions for the delta Bose gas and the Plancherel theorem it is possible to solve the above differential equation for any initial data. For delta initial data the solution can be simplified considerably so as to take the form of Theorem \ref{THMKPZMOMENTS}. As we observed in Section \ref{SecKPZFRED}, it is not possible to utilize the moments of the SHE to recover the distribution of, for instance, $z(t,x)$ for fixed $t$ and $x$. Nevertheless, Dotsenko \cite{Dot} and Calabrese-Le Doussal-Rosso \cite{CDR} reconstructed the known one-point distribution for $z(t,x)$ via the (mathematically non-rigorous) replica method using these moments.

\subsection{Be wise, discretize}

What is a possible mathematical interpretation for this replica method calculations of \cite{Dot,CDR}? To answer this question, we are drawn deeper into developing connections between the KPZ universality class and quantum integrable systems. The basic idea is that instead of working with the KPZ equation and delta Bose gas, we should first find an integrable discretization of the KPZ equation which converges to the equation under some scaling limit (such those in Section \ref{SECweak}). Second, we should identify some observes whose expectations (analogous to moments of the SHE) solve a quantum integrable system. Third, we should solve this system via the Bethe ansatz (developing the Plancherel theory as necessary) for general initial data. And fourth, we should utilize the resulting expectation formulas to compute distributional information about the model and take the limit to KPZ/SHE.

Steps one through three work for $q$-TASEP and ASEP (as well as a few other systems \cite{Cqmunu,CorPet,BCdiscrete}). So far step four has only been accomplished for some special types of initial data, including step (which we saw earlier is the discrete precursor to the fundamental solution to the SHE).

We will focus on this for $q$-TASEP and only briefly mention the case of ASEP which is treated analogously. Our aim is to provide an alternative proof (than that of Macdonald processes) to Theorem \ref{THMqTASEPmoments}.

\subsection{Duality between $q$-TASEP and the $q$-Boson process}\label{SECqduality}

\begin{figure}
\centering\includegraphics[width=12cm]{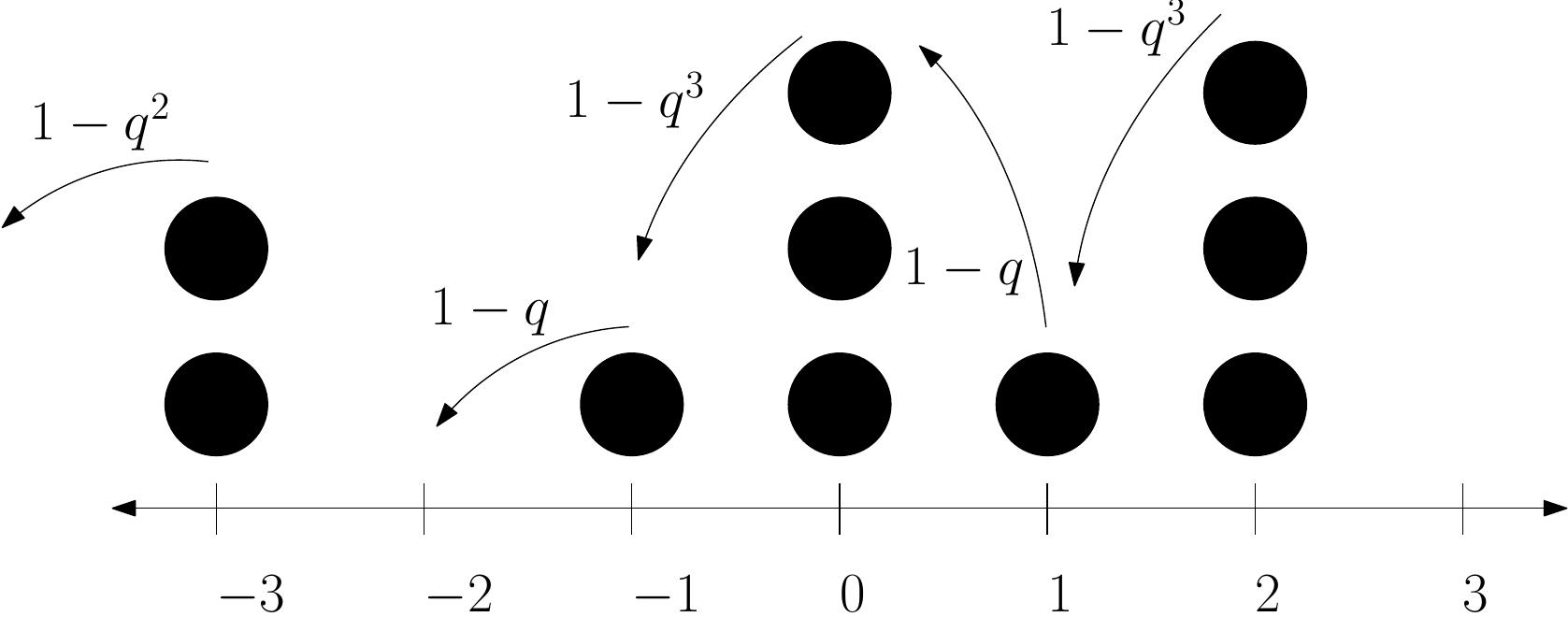}
\caption{The $q$-Boson process with $k=10$ particles at state $y_{-3}=2,y_{-1}=1,y_0=3,y_1=1,y_2=3$ and all other $y_j\equiv 0$. Equivalently, particles at ordered locations $\vec{n} = (2,2,2,1,0,0,0,-1,-3,-3)$ with $m=5$ clusters of sizes $\vec{c}= (3,1,3,1,2)$.}\label{qBoson}
\end{figure}

The $q$-Boson process was introduced by Sasamoto-Wadati \cite{SasWad} in 1998. It is a continuous time Markov process (a totally asymmetric zero range process) in which each site $j\in\Z$ has a non-negative number of particles $y_j$ sitting above it. In continuous time the top particle at each location $j$ jumps to the left by one site with a rate given by $1-q^{y_j}$. The process is illustrated in Figure \ref{qBoson} along with the notation $k$, $\vec{y}= \{y_j\}_{j\in \Z}$, $\vec{n}$, $\vec{c}$ and $m$. Assuming there are $k\geq 1$ particles in the system (particle count is preserved in time) it is also natural to record the state $\vec{y}$ as a vector $\vec{n} = (n_1\geq \cdots \geq n_k)$ of the ordered locations of the particles. Let $\vec{c} = (c_1,\ldots, c_m)$ represent the sizes of clusters in $\vec{n}$ and $m$ be the total number of such clusters.

The backward generator for the $q$-Boson process is
$$
\big(H f\big) (\vec{n}) = \sum_{i=1}^m \big(1-q^{c_i}\big) \big(f(\vec{n}_{c_1+\cdots + c_{i}}^{-}) - f(\vec{n})\big)
$$
where $f$ is a function of the ordered locations $\vec{n}$ and $\vec{n}_j^{-} = (n_1,\ldots, n_j-1,\ldots,n_k)$.

There is an obvious relationship between $q$-TASEP and the $q$-Boson process since the gaps of $q$-TASEP evolve according to the same zero range jumping rates as the $q$-Boson process. A less apparent, but quite useful (and simple to prove -- see \cite[Theorem 2.2]{BCS}) relationship is the {\it Markov duality} of these two processes. As $q$-TASEP is a discretization of the KPZ equation, this shows that the $q$-Boson process is a discretization of the delta Bose gas.

\begin{proposition}\label{Propduality}
For $q$-TASEP $x_n(t)$, $f(t,\vec{n}) := \EE\left[\prod_{j=1}^{k} q^{x_{n_j}(t)+n_j}\right]$ is the unique solution of
$$
\frac{\partial }{\partial t} f(t,\vec{n}) = \big(H f\big)(t,\vec{n}), \qquad \textrm{with} \qquad f(0,\vec{n}) = \EE\left[\prod_{j=1}^{k} q^{x_{n_j}(0)+n_j}\right].
$$
\end{proposition}

\subsection{Coordinate integrability of the $q$-Boson process}\label{SECcoord}
Define the {\it free generator} $\mathcal{L}$ via its action
$$
\big(\mathcal{L}u\big)(\vec{n}) = (1-q)\sum_{i=1}^{k} \big(\nabla_i u\big)(\vec{n})
$$
where $u:\Z^k\to \C$, $\big(\nabla f\big)(n) = f(n-1)-f(n)$, and $\nabla_i$ acts as $\nabla$ on coordinate $i$ of $u$. When all elements of $\vec{n}$ are unique (no clusters of $n$'s) the action of $\mathcal{L}$ matches that of $H$. The actions differ when clustering occurs. To repair this difference, we say that $u$ satisfies the {\it boundary conditions} if for all $1\leq i\leq k-1$
\begin{equation}\label{eqnBC}
\big(\nabla_i- q\nabla_{i+1}\big) u \big\vert_{n_i=n_{i+1}} = 0.
\end{equation}
The boundary conditions involve arguments $\vec{n}$ outside of the set of ordered $n_i$. The following result is proved as \cite[Proposition 2.7]{BCS}.

\begin{proposition}\label{Propsolving}
If $u:\R_{+}\times \Z^k\to \C$ satisfies the free evolution equation $\frac{\partial}{\partial t} u(t,\vec{n}) = \big(\mathcal{L} u\big)(t,\vec{n})$ and boundary conditions (\ref{eqnBC}), then its restriction to $\{n_1\geq \cdots \geq n_k\}$ satisfies the $q$-Boson process evolution equation $\frac{\partial}{\partial t} u(t,\vec{n}) = \big(Hu\big)(t,\vec{n})$.
\end{proposition}

Using Propositions \ref{Propduality} and \ref{Propsolving}  we can provide another proof of Theorem \ref{THMqTASEPmoments}. Let $u(t,\vec{n})$ be given by the  right-hand side of (\ref{EQNqTASEPmoments}). That $u$ satisfies the free evolution equation follows from the equality $$
\frac{\partial}{\partial t} \frac{e^{(q-1)tz}}{(1-z)^n}  = (1-q) \nabla \frac{e^{(q-1)tz}}{(1-z)^n}
$$
and Leibnitz rule. To check the boundary condition, observe that applying $\nabla_i-q\nabla_{i+1}$ to the integrand with $n_i=n_{i+1}$ results in a factor $z_i-qz_{i+1}$. This factor cancels the corresponding term in the denominator and allows the $z_i$ and $z_{i+1}$ contours to be freely deformed together thus showing that the remaining integral is zero by anti-symmetry. It remains to check initial data. Step initial data has $x_{n}(0)+n=0$ for all $n\geq 1$ and hence we must check that $u(0,\vec{n})\equiv \prod_{i=1}^{k} \mathbf{1}_{n_i\geq 1}$. This initial data is easily checked via residue calculus and comes from the poles of the product $\frac{1}{z_j}$ at zero.

The role that each term on the right-hand side of (\ref{EQNqTASEPmoments}) plays in solving the $q$-Boson process evolution equation suggests that one should look to generalize the $\frac{1}{z_i}$ product in order to study general initial data (which in turn corresponds to general $q$-TASEP initial data). In order to do this we develop the Plancherel theory necessary to diagonalize the $q$-Boson process generator via Bethe ansatz.

\subsubsection{Coordinate Bethe ansatz}
Consider a space $X$, an operator $L$ which acts on functions $f:X\to \C$, and an operator $B$ which acts on functions $g:X^2\to \C$. Let $\vec{x}=(x_1,\ldots, x_k)\in X^k$, $L_{i}$ act as $L$ on coordinate $i$ of functions $\Psi:X^k\to \C$, and $B_{i,i+1}$ act as $B$ on coordinates $i$ and $i+1$ of functions $\Psi:X^k\to \C$.

Algebraic eigenfunctions for an operator $\mathcal{L}$ acting on $\Psi:X^k\to \C$ as
$$
\big(\mathcal{L} \Psi\big)(\vec{x}) = \sum_{i=1}^{k} \big(L_i \Psi\big)(\vec{x})
$$
that satisfy boundary conditions
$$
B_{i,i+1} \Psi\big\vert_{x_i=x_{i+1}} = 0
$$
for $1\leq i\leq k-1$ can be diagonalized the following Bethe ansatz.
First, diagonalize the one dimensional operator $\big(L \psi_z\big)(x) = \lambda_z \psi_z(x)$ where $\psi_z:X\to C$ and $z\in \C$ indexes the eigenfunctions. Then consider linear combinations of products of these one dimensional eigenfunctions
$$
\Psi_{\vec{z}}(\vec{x}) := \sum_{\sigma\in S_k} A_{\sigma}(\vec{z}) \prod_{i=1}^{k} \psi_{z_{\sigma(i)}}(x_i).
$$
For arbitrary $\vec{z}\in \C^k$ and functions $A_{\sigma}(\vec{z})$ we must have
$$\big(\mathcal{L}\Psi_{\vec{z}}\big)(\vec{x}) = \Big(\sum_{i=1}^{k} \lambda_{z_i} \Big) \Psi_{\vec{z}}(\vec{x}).$$
Finally, choose
$$
A_{\sigma}(\vec{z}) := \sgn(\sigma) \prod_{k\geq A>B\geq 1} \frac{S(z_{\sigma(A)},z_{\sigma(B)})}{S(z_{A},z_{B})}\qquad\qquad \textrm{where} \qquad S(z_1,z_2) :=\frac{ B(\psi_{z_1}\otimes \psi_{z_2}\big)(x,x)}{\psi_{z_1}(x)\psi_{z_2}(x)}
$$
Then, for any $\vec{z}\in \C^k$ the corresponding $\Psi_{\vec{z}}(\vec{x})$ will be eigenfunctions of $\mathcal{L}$ which satisfy the boundary conditions. Since instead of working on a finite or periodic domain (often the setting of Bethe ansatz) we are working on $\Z$, there is no quantization of the spectrum (Bethe equations).

\subsubsection{Left and right eigenfunctions}
We apply Bethe ansatz to the $q$-Boson process Hamiltonian with $L=(1-q)\nabla$ and $B_{1,2} = \nabla_1 - q\nabla_2$ to compute the left eigenfunctions for $H$ (see \cite[Proposition 2.10]{BCPS}). While $H$ is not self-adjoint, it does enjoy a PT-invariance which immediately also yields right eigenfunctions.
\begin{proposition}\label{Propleftrighteigs}
For $\vec{z}\in \big(\C\setminus\{1\}\big)^k$ let
\begin{align*}
\Psi^{\ell}_{\vec{z}}(\vec{n}) &:= \sum_{\sigma\in S_k} \prod_{k\geq A>B\geq 1} \frac{z_{\sigma(A)}-q z_{\sigma(B)}}{z_{\sigma(A)}-z_{\sigma(B)}} \prod_{j=1}^{k} \frac{1}{(1-z_{\sigma(j)})^{n_j}},\\
\Psi^{r}_{\vec{z}}(\vec{n}) &:= \frac{1}{c_q(\vec{n})}\sum_{\sigma\in S_k} \prod_{k\geq A>B\geq 1} \frac{z_{\sigma(A)}-q^{-1} z_{\sigma(B)}}{z_{\sigma(A)}-z_{\sigma(B)}} \prod_{j=1}^{k} (1-z_{\sigma(j)})^{n_j},\\
\end{align*}
with $c_q(\vec{n}) = (-1)^k q^{-k(k-1)/2} (c_1)!_{q} (c_2)!_{q}\cdots$ (recall the $c_i$ are the cluster sizes for $\vec{n}$). Then
$$
H\Psi^{\ell}_{\vec{z}} = (1-q) (z_1+\cdots +z_k) \Psi^{\ell}_{\vec{z}}, \qquad H^t \Psi^{r}_{\vec{z}} = (1-q)(z_1+\cdots+ z_k) \Psi^{r}_{\vec{z}}
$$
where $H^t$ is the transpose of $H$.
\end{proposition}

\subsubsection{Direct and inverse Fourier type transforms}
Proposition \ref{Propleftrighteigs} gives algebraic eigenfunctions for $H$ corresponding to every $\vec{z}\in \big(\C\setminus\{1\}\big)^k$. That does not mean, however, that all of these eigenfunctions participate in diagonalizing $H$. For example, the Laplacian (acting in $x$ variables) has algebraic eigenfunctions $e^{zx}$ for all $z\in \C$. However, the decomposition of $L^2(\R)$ only involves those $z\in \I\R$. This fact is proved through the Plancherel theorem in Fourier analysis.

We define a direct and inverse Fourier type transform with respect to the $q$-Boson eigenfunctions. Let
\begin{align*}
\mathcal{W}^k &= \Big\{f:\{n_1\geq \cdots \geq n_k|n_j\in \Z\} \to \C \textrm{ of compact support}\Big\}\\
\mathcal{C}^k &= \C\big[(z_1-1)^{\pm 1},\ldots, (z_k-1)^{\pm 1}\big]^{S_k} = \textrm{symmetric Laurent polynomials in }(z_j-1), 1\leq j\leq k.
\end{align*}
The {\it direct transform} $\mathcal{F}:\mathcal{W}^k\to \mathcal{C}^k$ acts on $f\in \mathcal{W}^k$ as
$$
\big(\mathcal{F}f\big)(\vec{z}) := \sum_{n_1\geq \cdots \geq n_k} f(\vec{n}) \Psi^{r}_{\vec{z}}(\vec{n}) =: \big\langle f(\cdot),\Psi^r_{\vec{z}}(\cdot)\big\rangle_{\mathcal{W}}.
$$
The {\it inverse transform} $\mathcal{J}:\mathcal{C}^{k}\to \mathcal{W}^k$ acts on $G\in \mathcal{C}^{k}$ as
$$
\big(\mathcal{J} G\big)(\vec{n}) := \frac{(q-1)^{k} q^{-\frac{k(k-1)}{2}}}{(2\pi \I)^k k!} \oint \cdots \oint \det\left[\frac{1}{qw_i-w_j}\right]_{i,j=1}^{k} \prod_{j=1}^{k} \frac{w_j}{1-w_j} \Psi^{\ell}_{\vec{w}}(\vec{n}) G(\vec{w}) d\vec{w} =:\big\langle \Psi^{\ell}_{\cdot}(\vec{n}),G(\cdot)\big\rangle_{\mathcal{C}},
$$
where the contours are all along large circles around zero. Alternatively, the inverse transform can be put into a more familiar nested contour form (as we have seen before in Theorem \ref{THMqTASEPmoments})
$$
\big(\mathcal{J} G\big)(\vec{n}) = \frac{1}{(2\pi \I)^k} \oint \cdots \oint \prod_{1\leq A<B\leq k}\frac{z_A-z_B}{z_A-q z_B} \prod_{j=1}^{k} \frac{1}{(1-z_j)^{n_{j}+1}} G(\vec{z}) d\vec{z},
$$
where, for each $A\in \{1,\ldots,k\}$ the contour of integration of $z_A$ contains $1$, as well as $q$ times the contour of integration of $z_B$ for $B>A$, but does not contain $0$ (see Figure \ref{qContours}).

\subsubsection{Plancherel isomorphism theorem}
The following results are from \cite[Section 3]{BCPS}.

\begin{theorem}\label{TMPlan}
On the spaces $\mathcal{W}^k$ and $\mathcal{C}^k$, the operators $\mathcal{F}$ and $\mathcal{J}$ are mutual inverses of each other, 
and biorthogonal
\begin{align*}
\langle \Psi^{\ell}_{\cdot}(\vec{m}), \Psi^{r}_{\cdot}(\vec{n})\rangle_{\mathcal{C}} &= \mathbf{1}_{\vec{m}=\vec{n}},\\
\langle \Psi^{\ell}_{\vec{z}}(\cdot), \Psi^{r}_{\vec{w}}(\cdot)\rangle_{\mathcal{W}} &= \frac{1}{k!} \prod_{1\leq A\neq B\leq k} \frac{z_A-q z_B}{z_A-z_B} \prod_{j=1}^{k} \frac{1}{1-z_j} \det\big(\delta(z_i-w_j)\big)_{i,j=1}^{k}.
\end{align*}
\end{theorem}

This theorem diagonalizes the generator of the $q$-Boson process, proves completeness of the Bethe ansatz for it, and demonstrates remarkable biorthogonality properties of the eigenfunctions.

\subsubsection{Back to the $q$-Boson process}

An immediate corollary of Theorem \ref{TMPlan} (see \cite[Section 4]{BCPS}) is that for all initial data $f_0\in \mathcal{W}^k$, the unique solution to the $q$-Boson evolution equation
$$
\frac{\partial}{\partial t} f(t,\vec{n}) = \big(Hf\big)(t,\vec{n}),\qquad\textrm{with}\qquad f(0,\vec{n}) = f_0(\vec{n})
$$
equals
$$
f(t,\vec{n}) = \mathcal{J} \Big(e^{t(q-1)(z_1+\cdots +z_k)} \mathcal{F} f_0\Big)(\vec{n}) =  \frac{1}{(2\pi \I)^k} \oint \cdots \oint \prod_{1\leq A<B\leq k}\frac{z_A-z_B}{z_A-q z_B} \prod_{j=1}^{k} \frac{e^{t(q-1)z_j}}{(1-z_j)^{n_{j}+1}} \big(\mathcal{F}f_0\big)(\vec{z}) d\vec{z},
$$
where integration is along nested contours.

The limitation that $f_0\in \mathcal{W}^k$ can be relaxed (with some additional work). For instance, the above result can be extended to $f_0(\vec{n}) = \prod_{j=1}^{k} \mathbf{1}_{n_j\geq 1}$ which is the initial data corresponding to step initial data for $q$-TASEP (via the duality of Proposition \ref{Propduality}).

However, the computation of $\mathcal{F}f_0$ can still be difficult (it involves an infinite summation over weakly ordered $n_j$). If there is some $G\in \mathcal{C}^k$ for which $f_0=\mathcal{J} G$, then Theorem \ref{TMPlan} implies that
$$
\big(\mathcal{F} f_0\big)(\vec{z}) = G(\vec{z}).
$$
One easily checks that $G(\vec{z}) = q^{\frac{k(k-1)}{2}} \prod_{j=1}^{k} \frac{z_j-1}{z_j}$ yields $f_0(\vec{n}) = \prod_{j=1}^{k} \mathbf{1}_{n_j\geq 1}$. This (of course) agrees with our earlier solution to the $q$-Boson evolution equation.

\subsection{Algebraic integrability of the $q$-Boson system}

In 1998, Sasamoto-Wadati \cite{SasWad} first studied the $q$-Boson system (generalizing a similar system studied earlier in \cite{BBT,BIK}) via the language of algebraic Bethe ansatz.

The $q$-Boson algebra is generated by $B_j,B^{\dag}_j,N_j$, $1\leq j\leq M$ subject to the relations (usually $q$ would be replaced by $q^{-2}$, but the below parameterization is more convenient presently)
$$[B_i,B_j^{\dag}]=q^{N_i}\mathbf{1}_{i=j},\,[N_i,B_j]=-B_i\mathbf{1}_{i=j},\, [N_i,B_j^{\dag}]=B_i^{\dag}\mathbf{1}_{i=j}.$$
The period (size $M$ lattice) version of the $q$-Boson generator $H$ is the image of the $q$-Boson Hamiltonian
$$
\mathcal{H} = -(1-q)\sum_{j=1}^{M} \big(B_{j-1}^{\dag} - B_{j}^{\dag}\big) B_j
$$
under the representation in which $B_j,B_j^{\dag},N_j$ act on functions $f:(\Z_{\geq 0})^M \to \C$ as
$$
\big(B_j f\big)(\vec{y}) = \frac{1-q^{y_i}}{1-q} f(\cdots, y_j-1,\cdots), \qquad \big(B^{\dag}_j f\big)(\vec{y}) = f(\cdots, y_j+1,\cdots),\qquad \big(N_j f\big)(\vec{y}) = y_j f(\vec{y}).
$$
In \cite{SasWad}, $\mathcal{H}$ arises from the monodromy matrix of a quantum integrable system with trigonometric $R$-matrix, the same as in the XXZ and six-vertex model (as well as in ASEP). There are many questions which remain to be investigated regarding the use of the algebraic Bethe ansatz (of which this is an application) in producing interesting integrable probabilistic system.

In a different direction, the $q$-Boson generator $H$ also arises (see \cite[Lemma 6.1]{BCdiscrete}) from certain commutation relations for Macdonald first difference operators at Macdonald parameter $t=0$. Recall from Section \ref{SECdefmac} that $D^n_1$ is the Macdonald first difference operator acting on the variables $x_1,\ldots,x_n$.
\begin{proposition}
Assume the Macdonald parameter $t=0$, then
$$
\Big[\big(D_1^n)^k , p\Big] = (1-q^k) x_n \big(D_1^{n-1} - D_1^{n}\big) \big(D_1^n)^{k-1}
$$
where $p$ is the operator of multiplication by $(x_1+\cdots +x_n)$.
\end{proposition}

An immediate corollary of this is that for a symmetric, analytic function $F(x_1,\ldots, x_n)$,
$$
\tilde{f}(t,\vec{y}) = \begin{cases} 0,& \textrm{if at least one }y_{-j}>0, j\geq 0\\  e^{-tp}\big(D_1^1\big)^{y_1}\cdots \big(D_1^n\big)^{y_n} e^{tp} F(x_1,\cdots, x_n)\Big\vert_{x_1=\cdots=x_n=1},& \textrm{otherwise}\end{cases}
$$
solves the $q$-Boson evolution equation, in that $f(t,\vec{n}) = \tilde{f}(t,\vec{y}(\vec{n}))$ satisfies $\frac{\partial}{\partial t} f(t,\vec{n}) = \big(Hf\big)(t,\vec{n})$ where $\vec{y}= \vec{y}(\vec{n})$ is defined via $y_j = \big|\{i:n_i=j\}\big|$. Setting $F(x_1,\ldots,x_n)\equiv 1$ corresponds to step initial data for $q$-TASEP.

This provides one link between Macdonald processes and quantum integrable systems. Whether there is a deeper algebraic relationship between these two realms remains unclear.

\subsection{ASEP and beyond}\label{SECasepduality}

There is a parallel development for ASEP, as that explained above for $q$-TASEP. ASEP displays a non-trivial (self) duality \cite{Schutz,BCS} through which (recalling the notation of Section \ref{SECASEPmoments})
$$
f(t,\vec{y}):=\EE\left[\prod_{j=1}^{k} \tilde{Q}_{y_j}(t)\right]
$$
solves the ASEP backward equation (with $p$ and $q$ interchanged). This provides a route to checking the result of Theorem \ref{THMASEPmom}.

The ASEP generator can likewise be diagonalized via the Bethe ansatz, and a Plancherel theorem provides for the completeness of the ansatz and biorthogonality of the eigenfunctions (see \cite[Section 4 and 5]{BCS}). In fact, the ASEP and $q$-Boson Plancherel theorems are unified \cite{BCPS2} in terms of a theorem for the $(q,\mu,\nu)$-Boson process studied in \cite{Pov2,Cqmunu}. This Plancherel theorem also specializes to the general spin-$s$ XXZ model and to the six vertex model (on $\Z$).

These Plancherel theorems as well as the algebraic Bethe ansatz provide tools for further development of a theory of stochastic quantum integrable systems.

\end{document}